\documentclass{aa}
\usepackage[dvips]{graphicx}
\thesaurus{03(03.13.2; 03.20.7; 08.02.2; 08.22.1)}

\title{Spectral Analysis of Stellar Light Curves by Means of Neural 
Networks}

\author{R. Tagliaferri
           \inst{1} 
\and
        A. Ciaramella
           \inst{2}
\and
        L. Milano
           \inst{3} 
\and 
        F. Barone
           \inst{3} 
\and 
        G. Longo
           \inst{4}
}

\institute{
Dipartimento di Matematica ed Informatica, Universit\`a di
Salerno, via S. Allende, 84081 Baronissi (SA) Italia and INFM, unit\`a di 
Salerno, 84081 Baronissi (SA) Italia.
\and Universit\`a di Salerno, 84081 Baronissi (SA) Italia.
\and Dipartimento di Scienze Fisiche, Universit\`a di Napoli
``Federico II" and 
Istituto Nazionale di Fisica Nucleare, sez. Napoli,
Complesso Universitario di Monte Sant'Angelo, via Cintia, 80126 Napoli,
Italia.
\and Osservatorio Astronomico di Capodimonte, via Moiariello 16, 80131 Napoli, Italia.
}

\offprints{Roberto Tagliaferri - 
Dipartimento di Matematica ed Informatica, Universit\`a di
Salerno, via S. Allende, 84081 Baronissi (SA) Italia. \\
$^*$ This work was been partially supported by IIASS, by MURST 40\% and by the Italian Space Agency.}

\date{Received November 1997/Accepted 2 March 1999}

\begin{document}

\maketitle

\begin{abstract}

Periodicity analysis of unevenly collected data is a relevant issue in
several scientific fields. In astrophysics, for example, we have to find the
fundamental period of light or radial velocity curves which are unevenly
sampled observations of stars. Classical spectral analysis methods are
unsatisfactory to solve the problem. In this paper we present a neural
network based estimator system which performs well the frequency extraction
in unevenly sampled signals. It uses an unsupervised Hebbian nonlinear neural
algorithm to extract, from the interpolated signal, the principal components
which, in turn, are used by the MUSIC frequency estimator algorithm to
extract the frequencies. The neural network is tolerant to noise and works
well also with few points in the sequence. We benchmark the system on
synthetic and real signals with the Periodogram and with the Cramer-Rao
lower bound.

\end{abstract}

\section{Introduction}
The search for periodicities in time or spatial dependent signals is a topic
of the uttermost relevance in many fields of research, from geology
(stratigraphy, seismology, etc.; 
(Brescia et al. 1996) ) to astronomy (
Barone et al. 1994) where it finds wide application in the study of light
curves of variable stars, AGN's, etc.

The more sensitive instrumentation and observational techniques become, the
more frequently we find variable signals in time domain that previously were
believed to be constant. Research for possible periodicities in the signal
curves is then a natural consequence, when not an important issue. One of
the most relevant problems concerning the techniques of periodic signal
analysis is the way in which data are collected: many time series are
collected at unevenly sampling rate. We have two types of problems related
either to unknown fundamental period of the data, or their unknown multiple
periodicities. Typical cases, for instance in Astronomy, are the
determination of the fundamental period of eclipsing binaries both of light
and radial velocity curves, or the multiple periodicities determination of
ligth curves of pulsating stars. The difficulty arising from unevenly spaced
data is rather obvious and many attempts have been made to solve the problem
in a more or less satisfactory way. In this paper we will propose a new way
to approach the problem using neural networks, that seems to work
satisfactory well and seems to overcome most of the problems encountered
when dealing with unevenly sampled data.

\section{Spectral analysis and unevenly spaced data}

\subsection{Introduction}

In what follows, we assume $x$ to be a physical variable measured at
discrete times $t_{i}$. ${x(t_{i})}$ can be written as the sum of the signal 
$x_{s}$ and random errors $R$: $x_{i}=x(t_{i})=x_{s}(t_{i})+R(t_{i})$. The
problem we are dealing with is how to estimate fundamental frequencies which
may be present in the signal $x_{s}(t_{i})$ 
(Deeming 1975; Kay 1988; Marple 1987).

If $x$ is measured at uniform time steps (even sampling) 
(Horne \& Baliunas 1986; Scargle 1982) there are a lot of tools to
effectively solve the problem which are based on Fourier analysis 
(Kay 1988; Marple 1987; Oppennheim \& Schafer 1965). These methods, however,
are usually unreliable for unevenly sampled data. For instance, the typical
approach of resampling the data into an evenly sampled sequence, through
interpolation, introduces a strong amplification of the noise which affects
the effectiveness of all Fourier based techniques which are strongly
dependent on the noise level 
(Horowitz 1974).

There are other techniques used in specific areas 
(Ferraz-Mello 1981; Lomb 1976): however, none of them faces directly the
problem, so that they are not truly reliable. The most used tool for
periodicity analysis of evenly or unevenly sampled signals is the
Periodogram 
(Lomb 1976; Scargle 1982); therefore we will refer to it to evaluate our
system.

\subsection{Periodogram and its variations}

The Periodogram (P), is an estimator of the signal energy in the frequency
domain 
(Deeming 1975; Kay 1988; Marple 1987; Oppennheim \& Schafer 1965). It has
been extensively applied to pulsating star light curves, unevenly spaced,
but there are difficulties in its use, specially concerning with aliasing
effects.

\subsubsection{Scargle's Periodogram}

This tool is a variation of the classical P. It was introduced by J.D.
Scargle 
(Scargle 1982) for these reasons: 1) data from instrumental sampling are
often not equally spaced; 2) due to P inconsistency 
(Kay 1988; Marple 1987; Oppennheim \& Schafer 1965), we must introduce a
selection criterion for signal components.

In fact, in the case of even sampling, the classical P has a simple
statistic distribution: it is exponentially distributed for Gaussian noise.
In the uneven sampling case the distribution becomes very complex. However,
Scargle's P has the same distribution of the even case 
(Scargle 1982). Its definition is:

\begin{eqnarray}  \label{PScargle}
P_x(f) & = & \frac{1}{2} \frac{[\sum_{n=0}^{N-1} x(n)\cos 2\pi f(t_n-\tau)]^2}
{\sum_{n=0}^{N-1} \cos^2 2\pi f(t_n-\tau)} + \nonumber \\
& & \frac{[\sum_{n=0}^{N-1} x(n)\sin
2\pi f(t_n-\tau)]^2}{\sum_{n=0}^{N-1} \sin^2 2\pi f(t_n-\tau)}
\end{eqnarray}

\noindent
where 
\[
\tau=\frac{1}{4\pi f}\frac{\sum_{n=0}^{N-1} \sin 4\pi ft_n}{\sum_{n=0}^{N-1}
\cos 4\pi ft_n} 
\]

\smallskip
\noindent
and $\tau $ is a shift variable on the time axis, $f$ is the
frequency, $\{x\left( n\right) ,t_{n}\}$ is the observation series.

\subsubsection{Lomb's Periodogram}

This tool is another variation of the classical P and is similar to the
Scargle's P. It was introduced by Lomb 
(Lomb 1976) and we used the {\it Numerical Recipes in C} release (Numerical
Recipes in C 1988-1992).

Let us suppose to have $N$ points $x(n)$ and to compute mean and variance:

\begin{equation}
\bar{x}=\frac{1}{N}\sum_{n=1}^{N}x(n)\qquad \qquad \sigma ^{2}=\frac{1}{N-1}
\sum_{n=1}^{N}\left( x(n)-\bar{x}\right) ^{2}.  \label{eqc2.9}
\end{equation}

Therefore, the normalised Lomb's P (power spectra as function of an angular
frequency $\omega \equiv 2\pi f>0$) is defined as follows

\begin{eqnarray}
P_{N}(\omega ) & = & \frac{1}{2\sigma^{2}}\left[ \frac{[\sum_{n=0}^{N-1}\left(
x(n)-\bar{x}\right) \cos \omega (t_{n}-\tau )]^{2}}{\sum_{n=0}^{N-1}
\cos^{2}\omega (t_{n}-\tau )} \right] + \nonumber \\
 & & + \frac{1}{2\sigma^{2}} \left[ \frac{[\sum_{n=0}^{N-1}\left( x(n)-
\bar{x}\right) \sin \omega (t_{n}-\tau )]^{2}}{\sum_{n=0}^{N-1}\sin ^{2}\omega 
(t_{n}-\tau )} \right]  \label{PLomb}
\end{eqnarray}

\noindent
where $\tau $ is defined by the equation 
\[
\tan \left( 2\omega \tau \right) =\frac{\sum_{n=0}^{N-1}\sin 2\omega t_{n}}
{\sum_{n=0}^{N-1}\cos 2\omega t_{n}} 
\]

\smallskip 
\noindent
and $\tau $ is an offset, $\omega $ is the frequency, $\{x\left(n\right), 
t_{n}\}$ is the observation series. The horizontal lines in the figures 19, 22, 
25, 27, 32 and 34 correspond to the practical significance levels, as indicated 
in (Numerical Recipes in C 1988-1992).

\subsection{Modern spectral analysis}

Frequency estimation of narrow band signals in Gaussian noise is a problem
related to many fields 
(Kay 1988; Marple 1987). Since the classical methods of Fourier analysis
suffer from statistic and resolution problems, then newer techniques based
on the analysis of the signal autocorrelation matrix eigenvectors were
introduced 
(Kay 1988; Marple 1987).

\subsubsection{Spectral analysis with eigenvectors}

Let us assume to have a signal with p sinusoidal components (narrow band).
The p sinusoids are modelled as a stationary ergodic signal, and this is
possible only if the phases are assumed to be indipendent random variables
uniformly distributed in $[ 0,2\pi )$ 
(Kay 1988; Marple 1987). To estimate the frequencies we exploit the
properties of the signal autocorrelation matrix (a.m.) 
(Kay 1988; Marple 1987). The a.m. is the sum of the signal and the noise
matrices; the p principal eigenvectors of the signal matrix allow the
estimate of frequencies; the p principal eigenvectors of the signal matrix
are the same of the total matrix.

\section{PCA Neural Nets\label{section3}}

\subsection{Introduction}

Principal Component analysis (PCA) is a widely used technique in data
analysis. Mathematically, it is defined as follows: let ${\bf C}=E({\bf x}
{\bf x}^{T})$ be the covariance matrix of L-dimensional zero mean input data
vectors ${\bf x}$. The {\em i}-th principal component of ${\bf x}$ is
defined as ${\bf x}^{T}{\bf c}(i)$, where ${\bf c}(i)$ is the normalized
eigenvector of {\bf C} corresponding to the {\em i}-th largest eigenvalue 
$\lambda (i)$. The subspace spanned by the principal eigenvectors 
${\bf c}(1),\ldots ,{\bf c}(M),(M<L))$ is called the PCA subspace (of 
dimensionality M) 
(Oja et al. 1991; Oja et al. 1996). PCA's can be neurally realized in
various ways 
(Baldi \& Hornik 1989; Jutten \& Herault 1991; Oja 1982; Oja et al. 1991;
Plumbley 1993; Sanger 1989). The PCA neural network used by us is a one
layer feedforward neural network which is able to extract the principal
components of the stream of input vectors. Typically, Hebbian type learning
rules are used, based on the one unit learning algorithm originally proposed
by Oja 
(Oja 1982). Many different versions and extensions of this basic algorithm
have been proposed during the recent years; see 
(Karhunen \& Joutsensalo 1994; Karhunen \& Joutsensalo 1995; Oja et al.
1996; Sanger 1989).

\subsection{Linear, robust, nonlinear PCA Neural Nets}

The structure of the PCA neural network can be summarised as follows 
(Karhunen \& Joutsensalo 1994; Karhunen \& Joutsensalo 1995; Oja et al.
1996; Sanger 1989): there is one input layer, and one forward layer of
neurons totally connected to the inputs; during the learning phase there are
feedback links among neurons, that classify the network structure as either
hierarchical or symmetric. After the learning phase the network becomes
purely feedforward. The hierarchical case leads to the well known GHA
algorithm 
(Karhunen \& Joutsensalo 1995; Sanger 1989); in the symmetric case we have
the Oja's subspace network 
(Oja 1982).

PCA neural algorithms can be derived from optimisation problems, such as
variance maximization and representation error minimisation 
(Karhunen \& Joutsensalo 1994; Karhunen \& Joutsensalo 1995) so obtaining
nonlinear algorithms (and relative neural networks). These neural networks
have the same architecture of the linear ones: either hierarchical or
symmetric. These learning algorithms can be further classified in: robust
PCA algorithms and nonlinear PCA algorithms. We define robust a PCA algorithm 
When the objective function grows less than quadratically 
(Karhunen \& Joutsensalo 1994; Karhunen \& Joutsensalo 1995). The nonlinear
learning function appears at selected places only. In nonlinear PCA
algorithms all the outputs of the neurons are nonlinear function of the
responses.

\subsubsection{Robust PCA algorithms}

In the robust generalization of variance maximisation, the objective
function $f(t)$ is assumed to be a valid cost function 
(Karhunen \& Joutsensalo 1994; Karhunen \& Joutsensalo 1995), such as 
$\ln\cos (t)$ e $|t|$. This leads to the algorithm: 
\begin{eqnarray} \label{eq34}
{\bf w}_{k+1}(i) &=&{\bf w}_{k}(i)+\mu _{k}g(y_{k}(i)){\bf e}_{k}(i), \\
\qquad {\bf e}_{k}(i) &=&{\bf x}_{k}-\sum_{j=1}^{I(i)}y_{k}(j){\bf w}_{k}(j) 
\nonumber
\end{eqnarray}

In the hierarchical case we have $I(i)=i$. In the symmetric case $I(i)=M$,
the error vector ${\bf e}_k(i)$ becomes the same ${\bf e}_k$ for all the
neurons, and equation(\ref{eq34}) can be compactly written as: 
\begin{eqnarray}  \label{eq35}
{\bf W}_{k+1}={\bf W}_k+\mu{\bf e}_kg({\bf y}_k^T)
\end{eqnarray}

\noindent
where ${\bf y}={\bf W}^T_k{\bf x}$ is the instantaneous vector of neuron
responses. The learning function $g$, derivative of $f$, is applied
separately to each component of the argument vector.

The robust generalisation of the representation error problem 
(Karhunen \& Joutsensalo 1994; Karhunen \& Joutsensalo 1995), with 
$f(t)\le t^2$, leads to the stochastic gradient algorithm : 
\begin{eqnarray}  \label{eq37}
{\bf w}_{k+1}(i) & = & {\bf w}_k(i)+\mu ({\bf w}_k(i)^Tg({\bf e}_k(i))
{\bf x}_k + \\
                 & + & {\bf x}_k^T{\bf w}_k(i)g({\bf e}_k(i))) \nonumber
\end{eqnarray}

\noindent
This algorithm can be again considered in both hierarchical and symmetric
cases. In the symmetric case $I(i)=M$, the error vector is the same 
$({\bf e}_k)$ for all the weights ${\bf w}_k$. In the hierarchical case 
$I(i)=i$, equation(\ref{eq37}) gives the robust counterparts of principal 
eigenvectors ${\bf c}(i)$.

\subsubsection{Approximated Algorithms}

The first update term ${\bf w}_k(i)^Tg({\bf e}_k(i)){\bf x}_k$ in eq.(\ref
{eq37}) is proportional to the same vector ${\bf x}_k$ for all weights 
${\bf w}_k(i)$. Furthermore, we can assume that the error vector ${\bf e}_k$
should be relatively small after the initial convergence. Hence, we can
neglet the first term in equation(\ref{eq37}) and this leads to: 
\begin{equation}  \label{eq39}
{\bf w}_{k+1}(i)={\bf w}_k(i)+\mu {\bf x}_k^T y_k(i)g({\bf e}_k(i))
\end{equation}

\subsubsection{Nonlinear PCA Algorithms}

Let us consider now the nonlinear extensions of PCA algorithms. We can
obtain them in a heuristic way by requiring all neuron outputs to be always
nonlinear in the equation(\ref{eq34}) 
(Karhunen \& Joutsensalo 1994; Karhunen \& Joutsensalo 1995). This leads to:

\begin{eqnarray}
{\bf w}_{k+1}(i) & = & {\bf w}_{k}(i)+\mu g(y_{k}(i)){\bf b}_{k}(i),  
\label{eq310} \\
\qquad {\bf b}_{k}(i) &=& {\bf x}_{k}-\sum_{j=1}^{I(i)}g(y_{k}(j))
{\bf w}_{k}(j)\quad \forall i=1,\ldots ,p \nonumber
\end{eqnarray}

\section{Independent Component Analysis}

Independent Component Analysis (ICA) is a useful extension of PCA that was
developed in context with source or signal separation applications 
(Oja et al. 1996): instead of requiring that the coefficients of a linear
expansion of data vectors are uncorrelated, in ICA they must be mutually
independent or as independent as possible. This implies that second order
moments are not sufficient, but higher order statistics are needed in
determining ICA. This provides a more meaningful representation of data
than PCA. In current ICA methods based on PCA neural networks, the following
data model is usually assumed. The $L$-dimensional $k$-th data vector 
${\bf x}_{k}$ is of the form 
(Oja et al. 1996):

\begin{equation}
{\bf x}_{k}={\bf As}_{k}+{\bf n}_{k}=\sum_{i=1}^{M}s_{k}(i){\bf a}(i)+
{\bf n}_{k}  \label{eq311}
\end{equation}

where in the M-vector ${\bf s}_{k}=[s_{k}(1),\ldots ,s_{k}(M)]^{T}$, 
$s_{k}(i)$ denotes the $i$-th independent component (source signal) at time 
$k$, ${\bf A}=[{\bf a}(1),\ldots ,{\bf a}(M)]$ is a $L\times M$ {\em mixing
matrix} whose columns ${\bf a}(i)$ are the basis vectors of ICA, and 
${\bf n}_{k}$ denotes noise.

The source separation problem is now to find an $M\times L$ separating
matrix ${\bf B}$ so that the $M$-vector ${\bf y}_{k}={\bf Bx}_{k}$ is an
estimate ${\bf y}_{k}={\bf \hat{s}}_{k}$ of the original independent source
signal 
(Oja et al. 1996).

\subsection{Whitening}

Whitening is a linear transformation ${\bf A}$ such that, given a matrix 
${\bf C}$, we have ${\bf ACA}^{T}={\bf D}$ where ${\bf D}$ is a diagonal
matrix with positive elements 
(Kay 1988; Marple 1987).

Several separation algorithms utilise the fact that if the data vectors 
${\bf x}_{k}$ are first pre-processed by whitening them (i.e. 
$E(x_{k}x_{k}^{T})=I$ with $E(.)$ denoting the expectation), then the
separating matrix ${\bf B}$ becomes orthogonal (${\bf BB^{T}=I}$ see 
(Oja et al. 1996)).

Approximating contrast functions which are maximised for a separating matrix
have been introduced because the involved probability densities are unknown 
(Oja et al. 1996).

It can be shown that, for prewhitened input vectors, the simpler contrast
function given by the sum of kurtoses is maximised by a separating matrix 
${\bf B}$ 
(Oja et al. 1996).

However, we found that in our experiments the whitening was not as good as
we expected, because the estimated frequencies calculated for prewhitened
signals with the neural estimator (n.e.) were not too much accurate.

In fact we can pre-elaborate the signal, whitening it, and then we can apply
the n.e. . Otherwise we can apply the whitening and separate the signal in
independent components with the nonlinear neural algorithm of 
equation(\ref{eq310}) 
and then apply the n.e. to each of these components and estimate the
single frequencies separately.

The first method gives comparable or worse results than n.e. without
whitening. The second one gives worse results and is very expensive. When we
used the whitening in our n.e. the results were worse and more time
consuming than the ones obtained using the standard n.e. (i.e. without
whitening the signal). Experimental results are given in the following
sections. For these reasons whitening is not a suitable technique to improve
our n.e.\ .

\section{The neural network based estimator system}

The process for periodicity analysis can be divided in the following steps:

\par\noindent
{\bf - Preprocessing}
\par\noindent
We first interpolate the data with a simple linear fitting and then
calculate and subtract the average pattern to obtain zero mean process 
(Karhunen \& Joutsensalo 1994; Karhunen \& Joutsensalo 1995).

\par\noindent
{\bf - Neural computing}

The fundamental learning parameters are:
\par\noindent
{\bf 1)} the initial weight matrix;
\par\noindent
{\bf 2)} the number of neurons, that is the number of principal eigenvectors
that we need, equal to twice the number of signal periodicities (for real
signals);
\par\noindent
{\bf 3)} $\epsilon $, i.e. the threshold parameter for convergence ;
\par\noindent
{\bf 4)} $\alpha $, the nonlinear learning function parameter;
\par\noindent
{\bf 5)} $\mu $, that is the learning rate.

\par\noindent
We initialise the weight matrix ${\bf W}$ assigning the classical small
random values. Otherwise we can use the first patterns of the signal as the
columns of the matrix: experimental results show that the latter technique
speeds up the convergence of our neural estimator (n.e.). However, it cannot
be used with anomalously shaped signals, such as stratigraphic geological
signals.

Experimental results show that $\alpha $ can be fixed to : $1.$, $5.$, $10.$, 
$20.$, even if for symmetric networks a smaller value of $\alpha $ is
preferable for convergence reasons. Moreover, the learning rate $\mu $ can
be decreased during the learning phase, but we fixed it between $0.05$ and
$0.0001$ in our experiments.

We use a simple criterion to decide if the neural network has reached the
convergence: we calculate the distance between the weight matrix at step $%
k+1 $, ${\bf W}_{k+1}$, and the matrix at the previous step ${\bf W}_k$, and
if this distance is less than a fixed error threshold ($\epsilon$) we stop
the learning process.

We finally have the following general algorithm in which STEP 4 is one of
the neural learning algorithms seen above in section \ref{section3}:

\begin{itemize}
\itemindent=-0.5cm
\item[]{\bf STEP 1} Initialise the weight vectors ${\bf w}_{0}(i)\quad
\forall i=1,\ldots ,p$ with small random values, or with orthonormalised
signal patterns. Initialise the learning threshold $\epsilon $, the learning
rate $\mu $. Reset pattern counter $k=0$.

\item[]{\bf STEP 2} Input the k-th pattern ${\bf x}_{k}=[x(k),\ldots,
x(k+N+1)]$ where $N$ is the number of input components.

\item[]{\bf STEP 3} Calculate the output for each neuron $y(j) = 
{\bf w}^{T}(j){\bf x}_{i}\qquad \forall i=1,\ldots, p$.

\item[]{\bf STEP 4} Modify the weights ${\bf w}_{k+1}(i)={\bf w}_{k}(i)+\mu
_{k}g(y_{k}(i)){\bf e}_{k}(i)\qquad \forall i=1, \ldots, p$.

\item[]{\bf STEP 5} Calculate \begin{equation}
testnorma=\sqrt{\sum_{j=1}^{p}\sum_{i=1}^{N}
({\bf w}_{k+1}(ij)-{\bf w}_{k}(ij))^{2}}. \end{equation}

\item[]{\bf STEP 6} Convergence test: if $(testnorma < \epsilon )$ then goto 
{\bf STEP 8}.

\item[]{\bf STEP 7} $k=k+1$. Goto {\bf STEP 2}.

\item[]{\bf STEP 8} End.
\end{itemize}

\par\noindent
{\bf - Frequency estimator}
\par\noindent
We exploit the frequency estimator {\em Multiple Signal Classificator} (MUSIC). It
takes as input the weight matrix columns after the learning. The estimated
signal frequencies are obtained as the peak locations of the following
function 
(Kay 1988; Marple 1987): 
\begin{eqnarray}
P_{MUSIC} = \log(\frac{1}{1-\sum_{i=1}^M |{\bf e}_f^H{\bf w}(i)|^2}) & 
\end{eqnarray}

where ${\bf w}(i)$ is the $i-$th weight vector after learning, and ${\bf e}%
_f^H $ is the pure sinusoidal vector : ${\bf e}_f^H=[1,e_f^{j2\pi
f},\ldots,e_f^{j2\pi f(L-1)}]^H $.

When $f$ is the frequency of the $i-$th sinusoidal component, $f=f_i$, we
have ${\bf e} = {\bf e}_i $ and $P_{MUSIC} \to \infty$. In practice we have
a peak near and in correspondence of the component frequency. Estimates are
related to the highest peaks.

\section{Music and the Cramer-Rao Lower Bound}

In this section we show the relation between the Music estimator and the
Cramer-Rao bound following the notation and the conditions proposed by
Stoica and Nehorai in their paper (Stoica and Nehorai 1990).

\subsection{The model}

The problem under consideration is to determine the parameters of the
following model:

\begin{equation}
{\bf y}(t)=A({\bf \theta }){\bf x}(t)+{\bf e}(t)  \label{eq*.1}
\end{equation}

\noindent
where $\left\{ {\bf y}(t)\right\} \in C^{m\times 1}$ are the vectors of the
observed data, $\left\{ {\bf x}(t)\right\} \in C^{n\times 1}$ are the
unknown vectors and ${\bf e}(t)\in C^{m\times 1}$ is the added noise; the
matrix $A(\theta )\in C^{m\times n}$ and the vector $\theta $ are given by 
\begin{equation}
A({\bf \theta })=\left[ {\bf a}\left( \omega _{1}\right) ...{\bf a}\left(
\omega _{n}\right) \right] ;\qquad \qquad {\bf \theta }=\left[ \omega
_{1}...\omega _{n}\right]  \label{eq*.2}
\end{equation}

\noindent
where ${\bf a}\left( \omega \right) $ varies with the applications. Our aim
is to estimate the unknown parameters of ${\bf \theta }$. The dimension $n$
of ${\bf x}(t)$ is supposed to be known a priori and the estimate of the
parameters of ${\bf x}(t)$ is easy once ${\bf \theta }$ is known.

Now, we reformulate MUSIC to follow the above notation. The MUSIC estimate
is given by the position of the $n$ smallest values of the following
function: 
\begin{equation}
f\left( \omega \right) ={\bf a}^{\ast }\left( \omega \right) \hat{G}\hat{G}%
^{\ast }{\bf a}\left( \omega \right) ={\bf a}^{\ast }\left( \omega \right) %
\left[ I-\hat{S}\hat{S}^{\ast }\right] {\bf a}\left( \omega \right)
\label{eq*.3}
\end{equation}

From equation(\ref{eq*.3}) we can define the estimation error of a given
parameter. $\left\{ \hat{\omega}_{i}-\omega _{i}\right\} $ has (for big $N$)
an asintotic gaussian distribution, with $0$ mean and with the following
covariance matrix: 
\begin{equation}
C_{MU}=\frac{\sigma }{2n}\left( H\circ I \right)^{-1} 
Re \left\{ H\circ
\left( A^{\ast }UA\right)^{T} \right\} \left( H\circ I\right)^{-1}
\label{eq*.4}
\end{equation}

\noindent
where $Re \left( x\right) $ is the real part of $x$, where 
\begin{equation}
H=D^{\ast }GG^{\ast }D=D^{\ast }\left[ I-A\left( A^{\ast }A\right)
^{-1}A^{\ast }\right] D  \label{eq*.5}
\end{equation}

\noindent
and where $U$ is implicitly defined by: 
\begin{equation}
A^{\ast }UA=P^{-1}+\sigma P^{-1}\left( A^{\ast }A\right) ^{-1}P^{-1}
\label{eq*.6}
\end{equation}

\noindent
where $P$ is the covariance matrix of $x\left( t\right) $. The elements of
the diagonal of the matrix $C_{MU}$ are the variances of the estimation
error. On the other hand, the Cramer-Rao lower bound of the covariance
matrix of every estimator of ${\bf \theta }$, for large $N$, is given by: 
\begin{equation}
C_{CR}=\frac{\sigma }{2n}\left\{ Re \left[ H\circ P^{T}\right]
\right\}^{-1}.  \label{eq*.7}
\end{equation}

\noindent
Therefore the statistical efficiency can be defined with the condition that
P is diagonal as: 
\begin{equation}
\left[ C_{MU}\right] _{ii}\geq \left[ C_{CR}\right] _{ii}  \label{eq*.8}
\end{equation}

\noindent
where the equality is reached when $m$ increases if and only if 

\begin{equation}
{\bf a}^{\ast }\left( \omega \right) {\bf a}\left( \omega \right)
\longrightarrow \infty \qquad \qquad as 
\quad m\longrightarrow \infty .
\label{eq*.9}
\end{equation}

\noindent
For $P$ non-diagonal, $\left[ C_{MU}\right] _{ii}>\left[ C_{CR}\right]_{ii}$.
\noindent
To adapt the model used in the spectral analysis 
\begin{equation}
{\bf y}(k)=\sum_{i=1}^{p}A_{i}e^{j\omega _{i}k}+{\bf e}(k)\qquad \qquad
k=1,2,...,M  \label{eq*.10}
\end{equation}

\noindent
where $M$ is the total number of samples, to equation(\ref{eq*.3}) we make the
following transformations, after fixing an integer $m>p$: 
\begin{eqnarray}
{\bf y}(t) &=&\left[ y_{t}\quad ...\quad y_{t+m-1}\right]  \nonumber \\
{\bf a}\left( \omega \right) &=&\left[ 1\quad e^{j\omega }\quad e^{j\omega
\left( m-1\right) }\right]  \label{eq*.23} \\
{\bf x}(t) &=&\left[ A_{1}e^{j\omega _{1}t}\quad ...\quad A_{n}e^{j\omega
_{n}t}\right] \qquad t=1,...,M-m+1  \nonumber
\end{eqnarray}

\noindent
In this way our model satisfies the conditions of (Stoica and Nehorai 1990).
Moreover, equations(\ref{eq*.23}) depend on the choice of $m$ which influences
the minimum error variance.

\subsection{Comparison between PCA-MUSIC and the Cramer-Rao lower bound}

In this subsection we compare the n.e. method with the Cramer-Rao lower
bound, by varying the frequencies distance, the parameters $M$ and $m$ and
the noise variance.

From the experiments it derives that, fixed $M$ and $m$, by varying the
noise (white Gaussian) variance, the n.e. estimate is more accurate for
small values of the noise variance as shown in figures 1-3. For $\Delta \omega $
small, the noise variance is far from the bound. By increasing $m$ the
estimate improves, but there is a sensitivity to the noise (figures 4-6). By
varying $M$, there is a sensitivity of the estimator to the number of points
and to $m$ (figures 7-8). In fact, if we have a quite large number of points we
reach the bound as illustrated in figures 9-10.

\begin{figure}
\includegraphics[width=8cm , height=7cm]{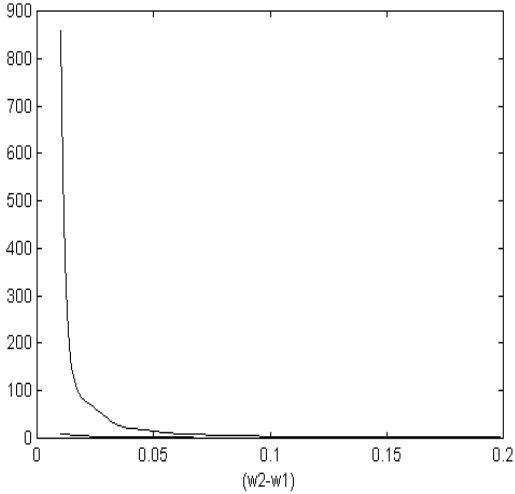}
\caption[]{CRB and standard deviation of n.e. estimates; abscissa is 
the distance between the frequencies $\omega_{2}$ and $\omega_{1}$. \\
CRB (down); standard deviation of n.e. (up) with $m=5$, $\sigma=0.5$ and 
$M=50$.}
\end{figure}

\begin{figure}
\vbox{
\hbox{
\includegraphics[width=8cm , height=7cm]{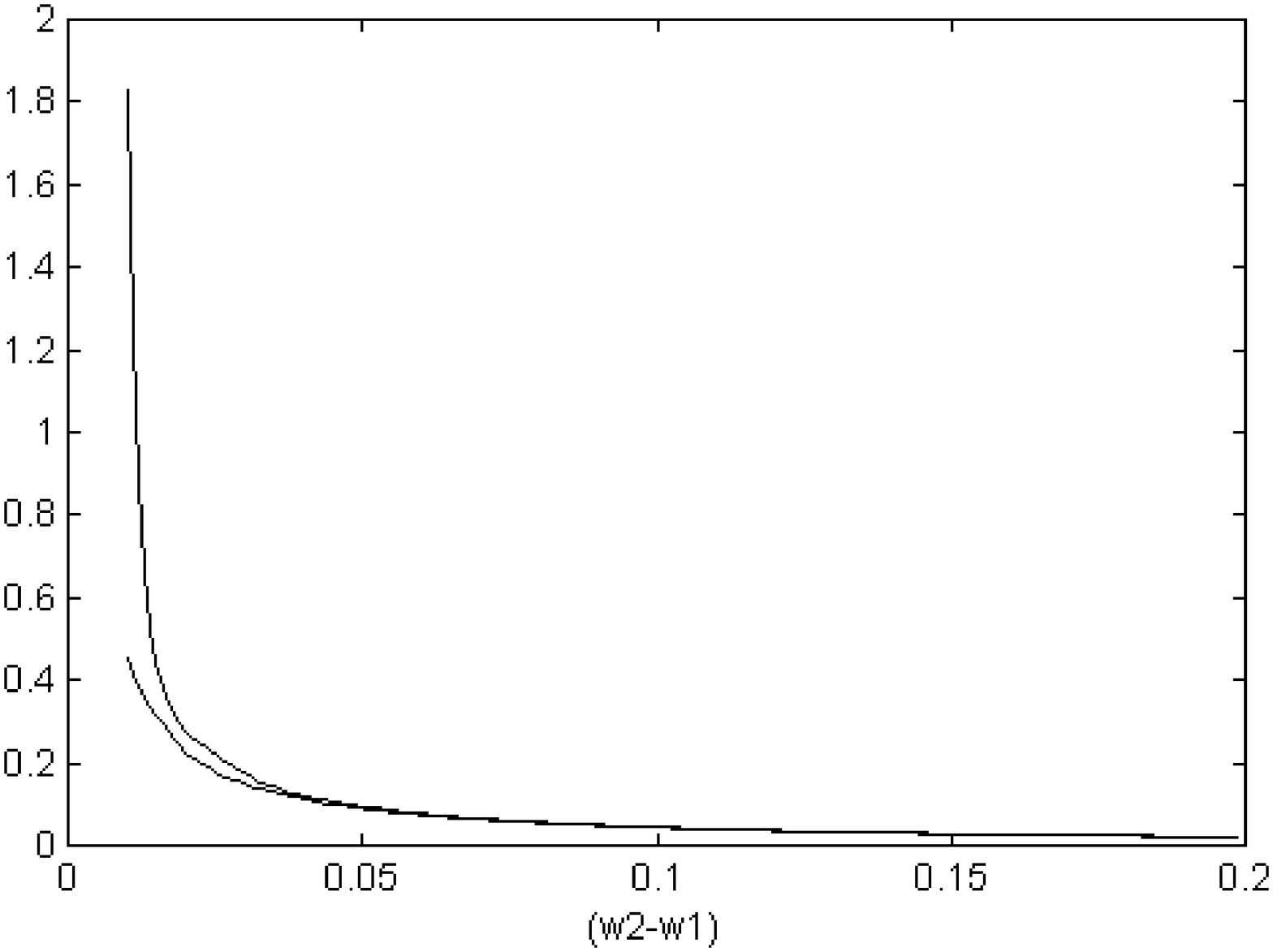}

}}
\caption[]{CRB and standard deviation of n.e. estimates; abscissa is 
the distance between the frequencies $\omega_{2}$ and $\omega_{1}$. \\
CRB (down); standard deviation of n.e. (up) with $m=5$, $\sigma=0.001$ 
and $M=50$.}
\end{figure}

\begin{figure}
\vbox{
\hbox{
\includegraphics[width=8cm , height=7cm]{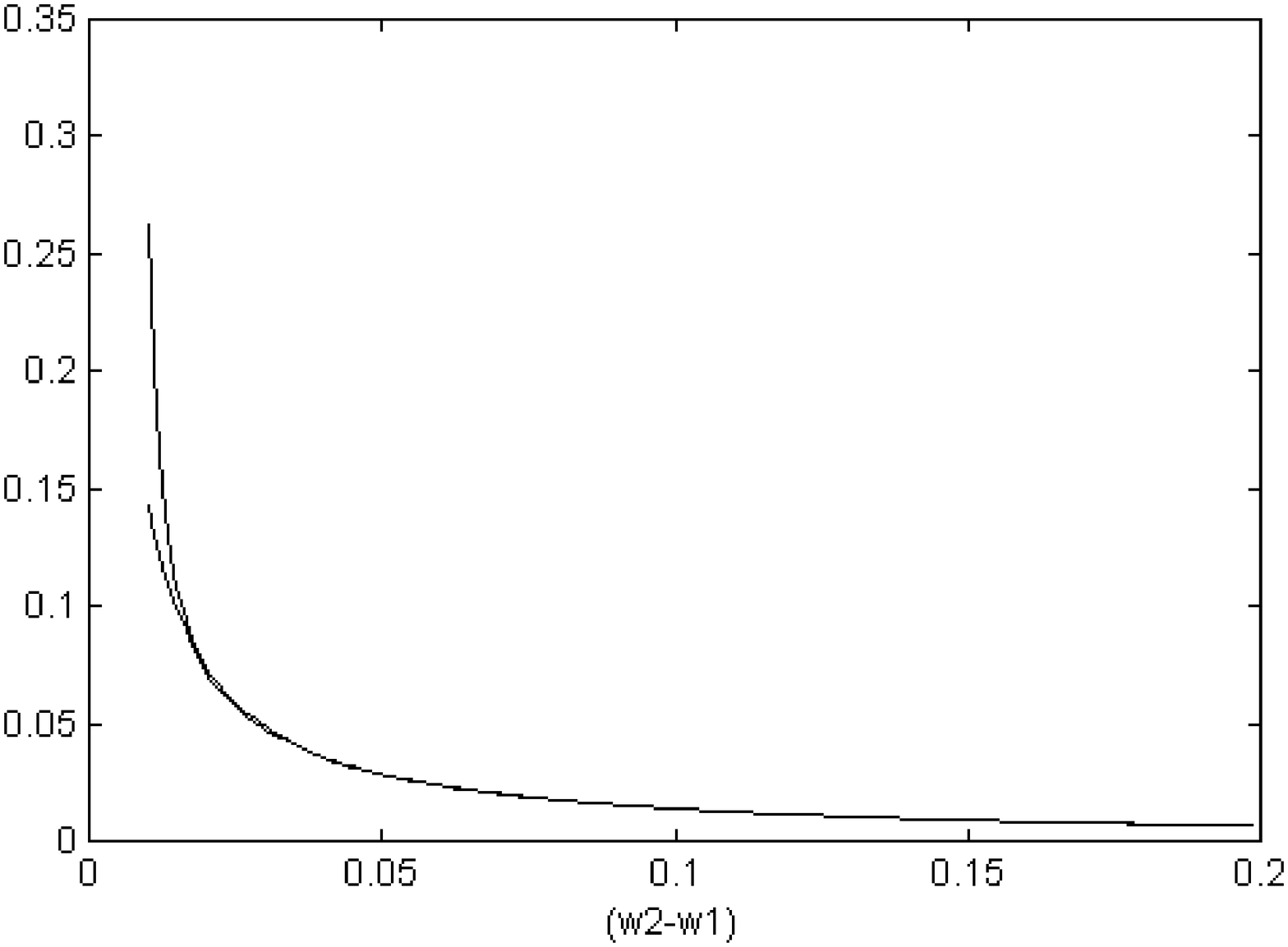}
}}
\caption[]{CRB and standard deviation of n.e. estimates; abscissa is 
the distance between the frequencies $\omega_{2}$ and $\omega_{1}$. \\
CRB (down); standard deviation of n.e. (up) with $m=5$, $\sigma=0.0001$ and 
$M=50$.}
\end{figure}

\begin{figure}
\vbox{
\hbox{
\includegraphics[width=8cm , height=7cm]{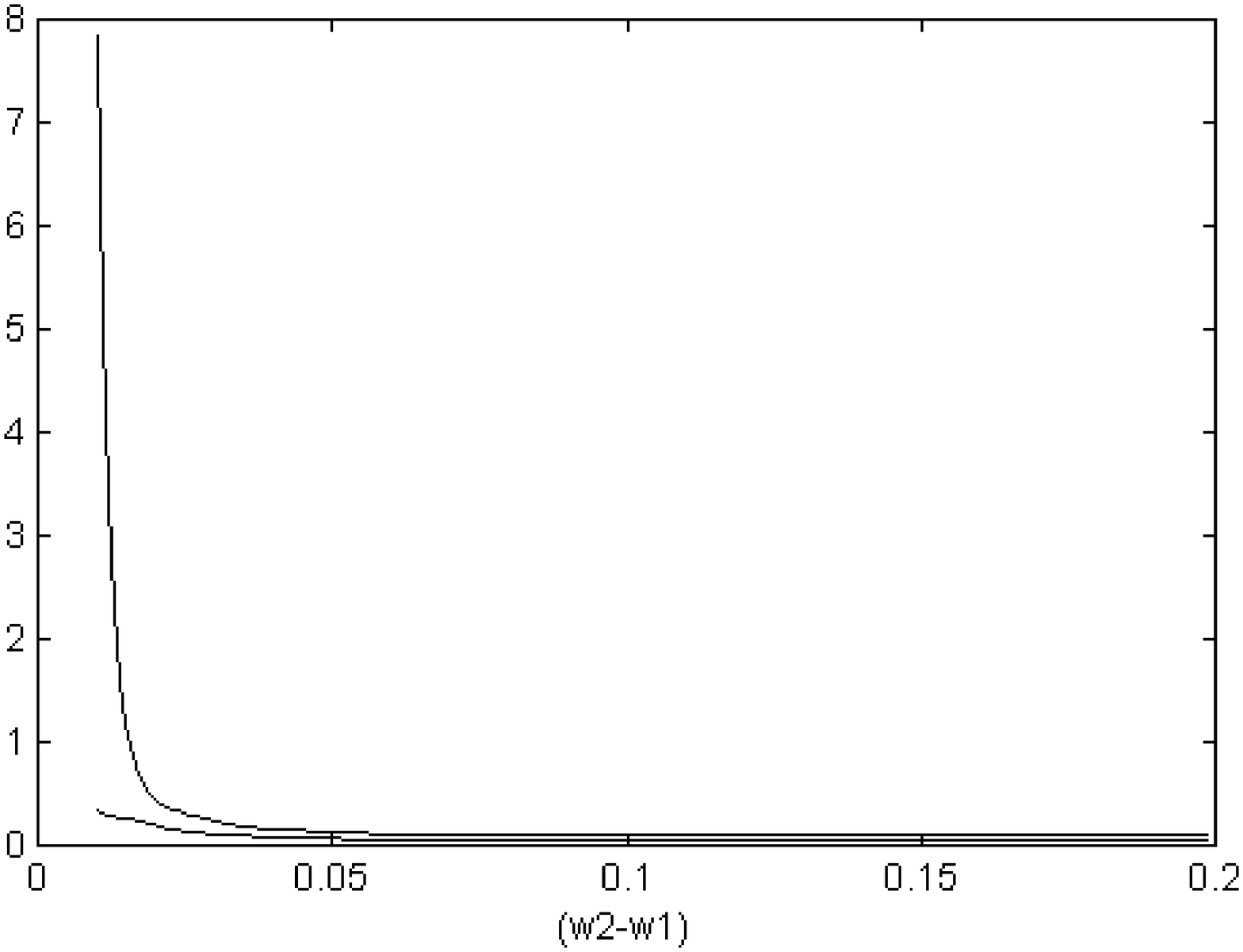}

}}
\caption[]{CRB and standard deviation of n.e. estimates; abscissa is 
the distance between the frequencies $\omega_{2}$ and $\omega_{1}$. \\
CRB (down); standard deviation of n.e. (up) with $m=20$, $\sigma=0.5$ and 
$M=50$.}
\end{figure}

\begin{figure}
\vbox{
\hbox{
\includegraphics[width=8cm , height=7cm]{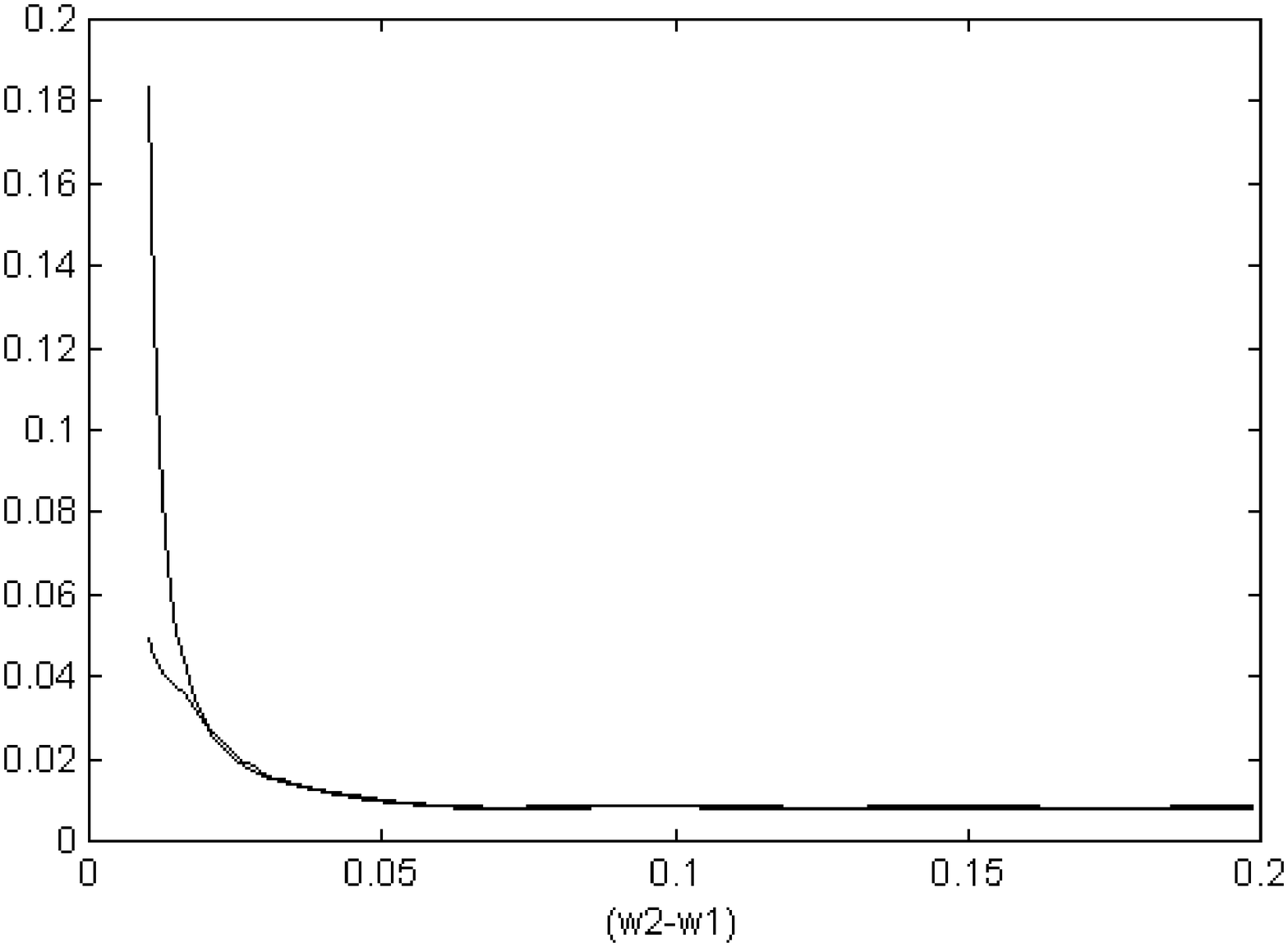}

}}
\caption[]{CRB and standard deviation of n.e. estimates; abscissa is 
the distance between the frequencies $\omega_{2}$ and $\omega_{1}$. \\
CRB (down); standard deviation of n.e. (up) with $m=20$, $\sigma=0.01$ and 
$M=50$.}
\end{figure}

\begin{figure}
\vbox{
\hbox{
\includegraphics[width=8cm , height=7cm]{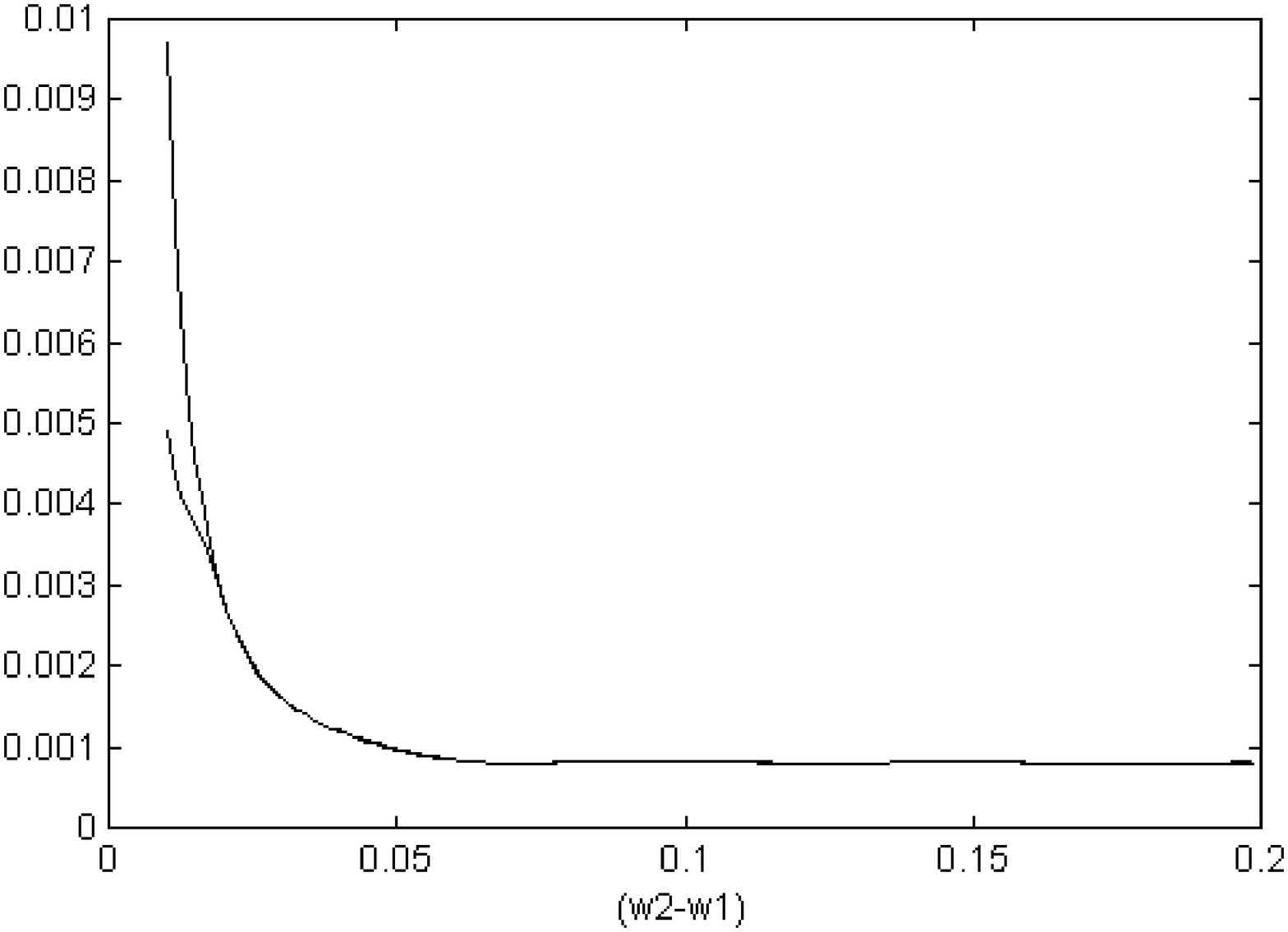}

}}
\caption[]{CRB and standard deviation of n.e. estimates; abscissa is 
the distance between the frequencies $\omega_{2}$ and $\omega_{1}$. \\
CRB (down); standard deviation of n.e. (up) with $m=20$, $\sigma=0.0001$ and 
$M=50$.}
\end{figure}

\begin{figure}
\vbox{
\hbox{
\includegraphics[width=8cm , height=7cm]{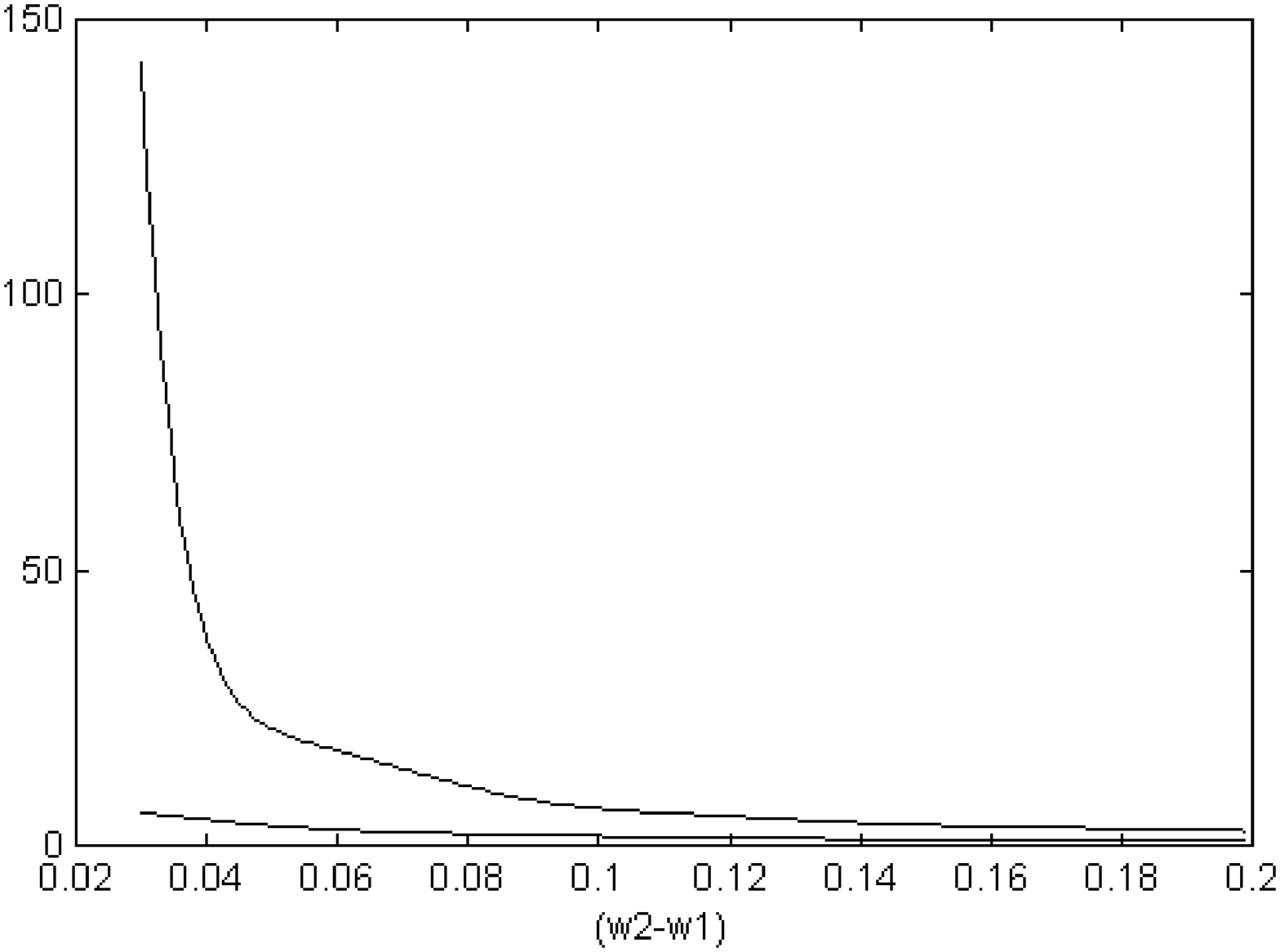}

}}
\caption[]{CRB and standard deviation of n.e. estimates; abscissa is 
the distance between the frequencies $\omega_{2}$ and $\omega_{1}$. \\
CRB (down); standard deviation of n.e. (up) with $m=5$, $\sigma=0.01$ and 
$M=20$.}
\end{figure}

\begin{figure}
\vbox{
\hbox{
\includegraphics[width=8cm , height=7cm]{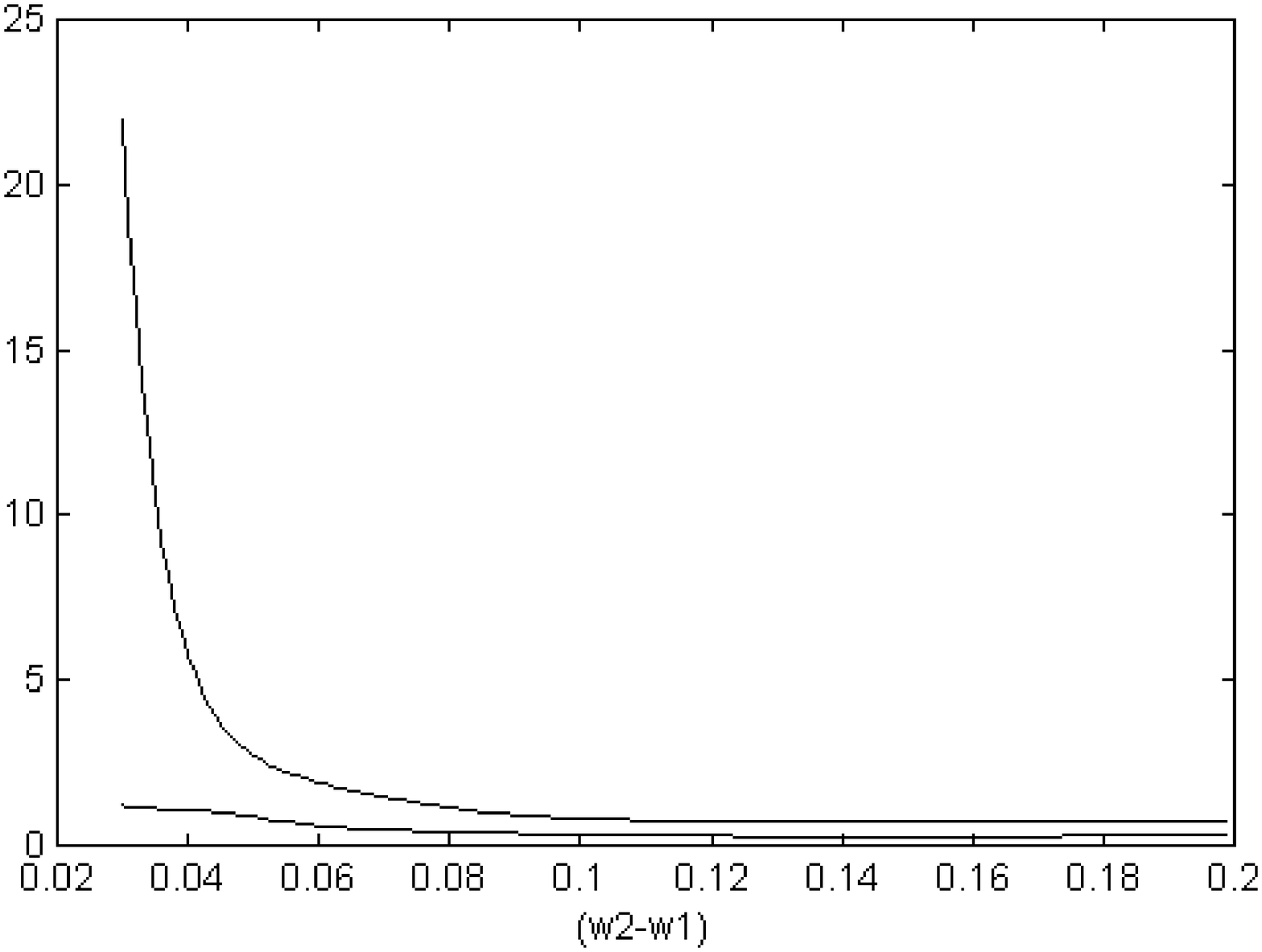}

}}
\caption[]{CRB and standard deviation of n.e. estimates; abscissa is 
the distance between the frequencies $\omega_{2}$ and $\omega_{1}$. \\
CRB (down); standard deviation of n.e. (up) with $m=5$, $\sigma=0.01$ and 
$M=20$.}
\end{figure}

\begin{figure}
\vbox{
\hbox{
\includegraphics[width=8cm , height=7cm]{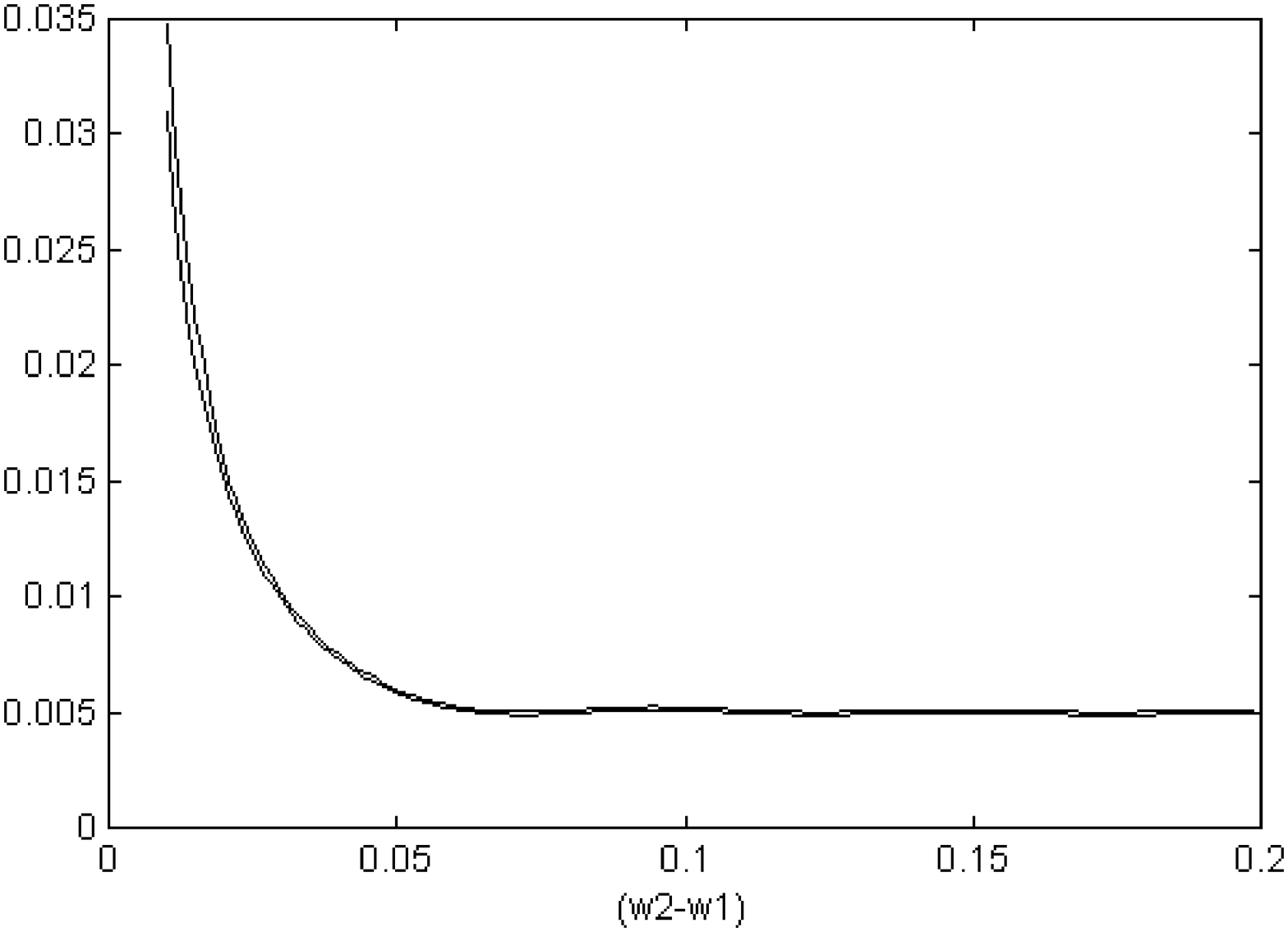}

}}
\caption[]{CRB and standard deviation of n.e. estimates; abscissa is 
the distance between the frequencies $\omega_{2}$ and $\omega_{1}$. \\
CRB (down); standard deviation of n.e. (up) with $m=20$, $\sigma=0.01$ and 
$M=100$.}
\end{figure}

\begin{figure}
\vbox{
\hbox{
\includegraphics[width=8cm , height=7cm]{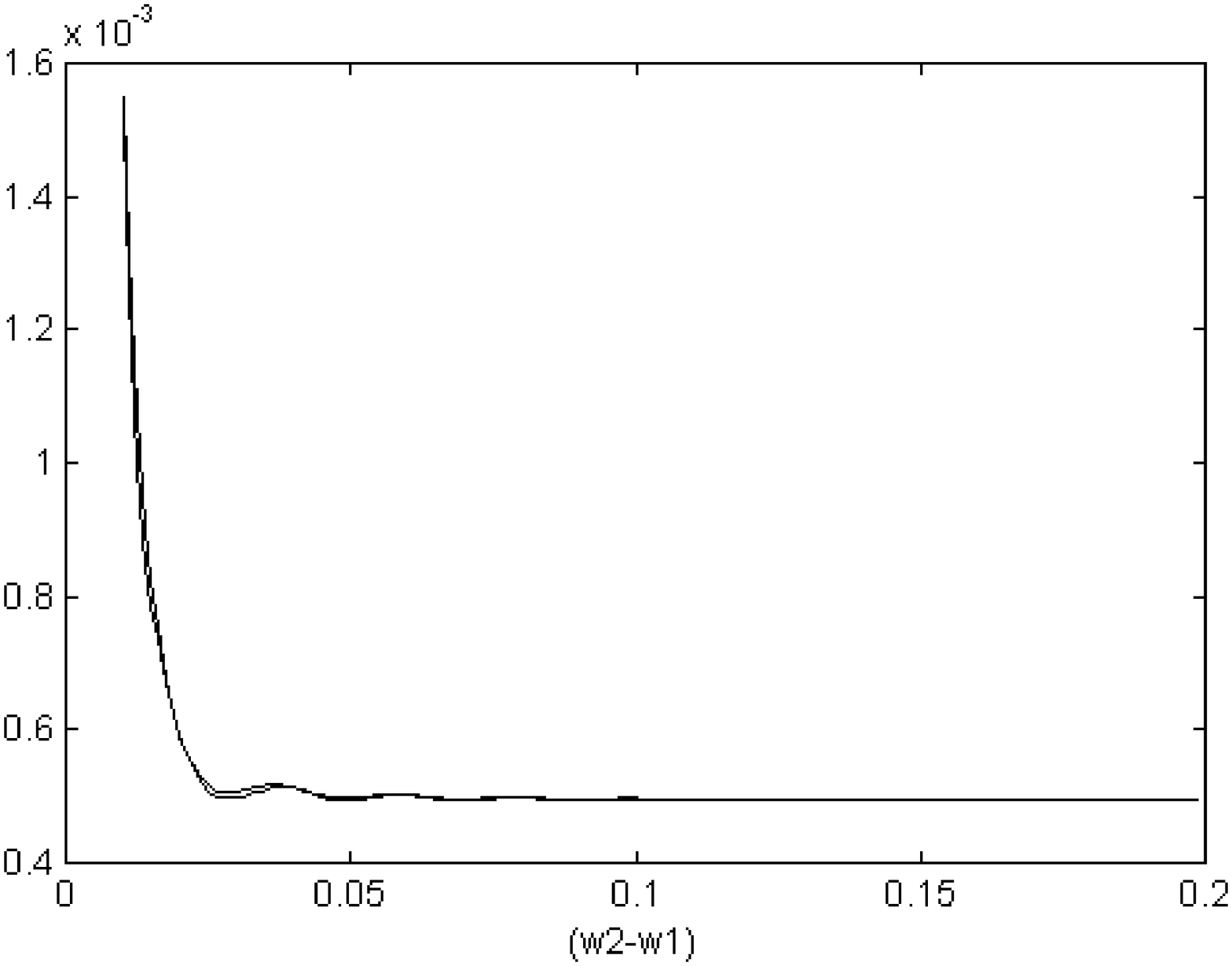}

}}
\caption[]{CRB and standard deviation of n.e. estimates; abscissa is 
the distance between the frequencies $\omega_{2}$ and $\omega_{1}$. \\
CRB (down); standard deviation of n.e. (up) with $m=50$, $\sigma=0.001$ and 
$M=100$.}
\end{figure}

Therefore, the n.e. estimate depends on both the increase of $m$ and the
number of points in the input sequence. Increasing the number of points, we
improve the estimate and the error approximates the Cramer-Rao bound. On the
other hand, for noise variances very small, the estimate reaches a very good
performance. Finally, we see that in all the experiments shown in the
figures we reach the bound with a good approximation, and we can conclude
that the n.e.\ method is statistically efficient.

\section{Experimental results}

\subsection{Introduction}

In this section we show the performance of the neural based estimator system
using artificial and real data. The artificial data are generated following
the literature 
(Kay 1988; Marple 1987) and are noisy sinusoidal mixtures. These are used to
select the neural models for the next phases and to compare the n.e. with
P's, by using Montecarlo methods to generate samples. Real data, instead,
come from astrophysics: in fact, real signals are light curves of Cepheids
and a light curve in the Johnson's system.

In the sections \ref{realistic data} and \ref{real data}, we use an
extension of Music to directly include unevenly sampled data without using
the interpolation step of the previous algorithm in the following way:

\begin{equation}
P_{MUSIC}^{\prime }=\frac{1}{M-\sum_{i=1}^{p}|{\bf e}_{f}^{H}{\bf w}(i)|^{2}}
\end{equation}

\noindent
where $p$ is the frequency number, ${\bf w}(i)$ is the $i-$th weight vector
of the PCA neural network after the learning, and ${\bf e}_{f}^{H}$ is the
sinusoidal vector: ${\bf e}_{f}^{H}=[1,e_{f}^{j2\pi ft_{0}},\ldots
,e_{f}^{j2\pi ft_{(L-1)}}]^{H}$ where $\left\{ t_{0},t_{1},...,t_{\left(
L-1\right) }\right\} $ are the first $L$ components of the temporal
coordinates of the uneven signal.

Furthermore, to optimise the performance of the PCA neural networks, we stop
the learning process when $\sum_{i=1}^{p}|{\bf e}_{f}^{H}{\bf w}%
(i)|^{2}>M\qquad \forall f$, so avoiding overfitting problems.

\subsection{Models selection\label{model selection}}

In this section we use synthetic data to select the neural networks used in
the next experiments. In this case, the uneven sampling is obtained by
randomly deleting a fixed number of points from the synthetic
sinusoid-mixtures: this is a widely used technique in the specialised
literature 
(Horne \& Baliunas 1986).

The experiments are organised in this way. First of all, we use synthetic
unevenly sampled signals to compare the different neural algorithms in the
neural estimator (n.e.) with the Scargle's P.

For this type of experiments, we realise a statistical test using five
synthetic signals. Each one is composed by the sum of five sinusoids of
randomly chosen frequencies in $[0,0.5]$ and randomly chosen phases in $%
[0,2\pi ]$ 
(Kay 1988; Karhunen \& Joutsensalo 1994; Marple 1987), added to white random
noise of fixed variance. We take $200$ samples of each signal and randomly
discard $50\%$ of them ($100$ points), getting an uneven sampling 
(Horne \& Baliunas 1986). In this way we have several degree of randomness:
frequencies, phases, noise, deleted points.

After this, we interpolate the signal and evaluate the P and the n.e. system
with the following neural algorithms: robust algorithm in equation(\ref{eq34}) 
in
the hierarchical and symmetric case; nonlinear algorithm in 
equation(\ref{eq310})
in the hierarchical and symmetric case. Each of these is used with two
nonlinear learning functions: $g_1(t)=\tanh(\alpha t)$ and $%
g_2(t)=sgn(t)\log (1+\alpha|t|)$. Therefore we have eight different neural
algorithms to evaluate.

We chose these algorithms after we made several experiments involving all
the neural algorithms presented in section \ref{section3}, with several
learning functions, and we verified that the behaviour of the algorithms and
learning functions cited above was the same or better than the others. So we
restricted the range of algorithms to better show the most relevant features
of the test.

We evaluated the average differences between target and estimated
frequencies. This was repeated for the five signals and then for each
algorithm we made the average evaluation of the single results over the five
signals. The less this averages were, the greatest the accuracy was.

We also calculated the average of the number of epochs and CPU time for
convergence. We compare this with the behaviour of P.

Common signals parameters are: number of frequencies $=5$, variance noise $%
=0.5$, number of sampled points $=200$, number of deleted points $=100$.
\par\noindent
Signal 1: frequencies=$0.03, 0.19, 0.25, 0.33, 0.46 \quad 1/s $
\par\noindent
Signal 2: frequencies=$0.02, 0.11, 0.20, 0.33, 0.41 \quad 1/s $
\par\noindent
Signal 3: frequencies=$0.34, 0.29, 0.48, 0.42, 0.04 \quad 1/s $
\par\noindent
Signal 4: frequencies=$0.32, 0.20, 0.45, 0.38, 0.13 \quad 1/s $
\par\noindent
Signal 5: frequencies=$0.02, 0.37, 0.16, 0.49, 0.31 \quad 1/s $
\par\noindent
Neural parameters: $\alpha =10.0$; $\mu =0.0001$; $\epsilon =0.001$; number
of points in each pattern $N=110$ (these are used for almost all the neural
algorithms; however, for a few of them a little variation of some parameters
is required to achieve convergence).
\par\noindent
Scargle parameters: $Tapering=30\%$, $p_0=0.01$.
\par\noindent
Results are shown in Table~\ref{table61}:

\par\noindent
\begin{table*}[tbp]
\caption[]{Performance evaluation of n.e. algorithms and P on 
synthetic signals.}
\label{table61}

\begin{flushleft}
\begin{tabular}{|c|c|c|c|c|c|c|c|c|}
\hline
& \multicolumn{6}{|c|}{} & \multicolumn{2}{|c|}{} \\ 
& \multicolumn{6}{|c|}{average normalised differences} & 
\multicolumn{2}{|c|}{} \\ 
& \multicolumn{6}{|c|}{} & \multicolumn{2}{|c|}{} \\ \hline
&  &  &  &  &  &  &  &  \\ 
Algorithm & sig1 & sig2 & sig3 & sig4 & sig5 & TOT & average & average \\ 
&  &  &  &  &  &  & n. epochs & time \\ 
&  &  &  &  &  &  &  &  \\ \hline\hline
&  &  &  &  &  &  &  &  \\ 
1. eq.(\ref{eq34}) hierarc.+$g1$ & 0.000 & 0.002 & 0.004 & 0.000 & 0.004 & 
0.0020 & 898.4 & 189.2 s \\ 
2. eq.(\ref{eq34}) hierarc.+$g2$ & 0.000 & 0.002 & 0.004 & 0.000 & 0.004 & 
0.0020 & 667.2 & 105.2 s \\ 
3. eq.(\ref{eq310}) hierarc.+$g1$ & 0.000 & 0.002 & 0.005 & 0.000 & 0.004 & 
0.0022 & 5616.2 & 1367.4 s \\ 
4. eq.(\ref{eq310}) hierarc.+$g2$ & 0.000 & 0.002 & 0.005 & 0.000 & 0.004 & 
0.0022 & 3428.4 & 1033.4 s \\ 
5. eq.(\ref{eq34}) symmetr.+$g1$ & 0.000 & 0.002 & 0.002 & 0.000 & 0.004 & 
0.0016 & 814.0 & 100.2 s \\ 
6. eq.(\ref{eq34}) symmetr.+$g2$ & 0.000 & 0.002 & 0.004 & 0.002 & 0.004 & 
0.0024 & 855.2 & 124.4 s \\ 
7. eq.(\ref{eq310}) symmetr.+$g1$ & 0.000 & 0.002 & 0.004 & 0.002 & 0.004 & 
0.0024 & 6858.2 & 1185 s \\ 
8. eq.(\ref{eq310}) symmetr.+$g2$ & 0.000 & 0.002 & 0.004 & 0.002 & 0.004 & 
0.0024 & 3121.8 & 675.8 s \\ 
Periodogram & 0.004 & 0.000 & 0.002 & 0.004 & 0.004 & 0.0028 &  & 22.2 s \\ 
&  &  &  &  &  &  &  &  \\ \hline
\end{tabular}
\end{flushleft}
\end{table*}

We have to spend few words about the differences of behaviour among the
neural algorithms elicited by the experiments. Nonlinear algorithms are more
complex than robust ones; they are relatively slower in converging, with
higher probability to be caught in local minima, so their estimates results
are sometimes not reliable. So we restrict our choice to robust models.
Moreover, symmetric models require more effort in finding the right
parameters to achieve convergence than the hierarchical ones. The
performance, however, are comparable.

From Table~\ref{table61} we can see that the best neural algorithm for our
aim is the n.5 in Table~\ref{table61} (equation(\ref{eq34}) in the symmetric 
case
with learning function $g_1(t)=\tanh(\alpha t)$).

However, this algorithm requires much more efforts in finding the right
parameters for the convergence than the algorithm n.2 from the same table
(equation(\ref{eq34}) in the hierarchical case with learning function 
$g_{2}(t)=sgn(t)\log (1+\alpha |t|)$), which has performance comparable with
it.

For this reason, in the following experiments when we present the neural
algorithm, it is algorithm n.2.

We show, as an example, in figures 11-13 the estimate result of the n.e.
algorithm and P on signal n.1.

\begin{figure}
\vbox{
\hbox{
\includegraphics[width=8cm , height=7cm]{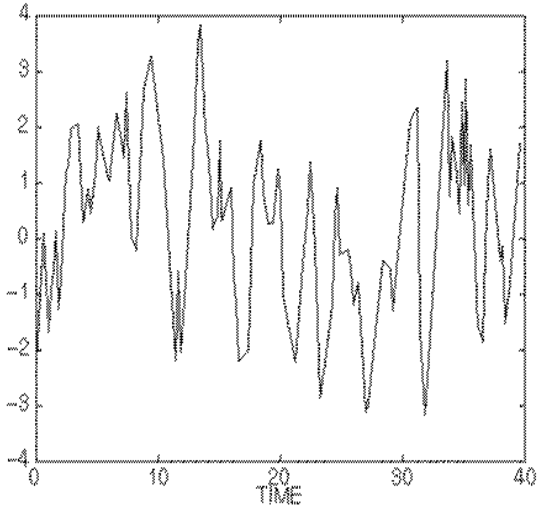}

}}
\caption[]{Synthetic Signal.}
\end{figure}

\begin{figure}
\vbox{
\hbox{
\includegraphics[width=8cm , height=7cm]{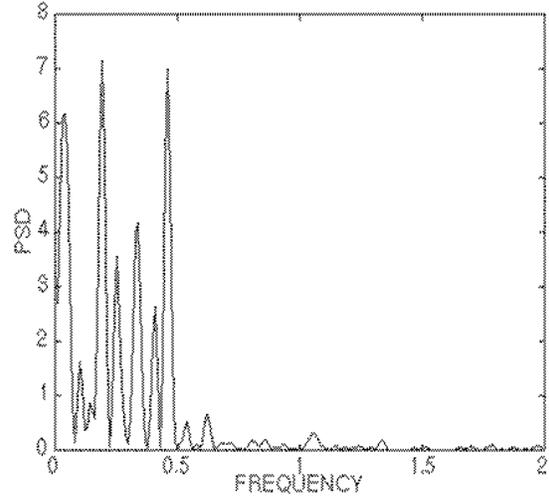}

}}
\caption[]{P estimate.}
\end{figure}

\begin{figure}
\vbox{
\hbox{
\includegraphics[width=8cm , height=7cm]{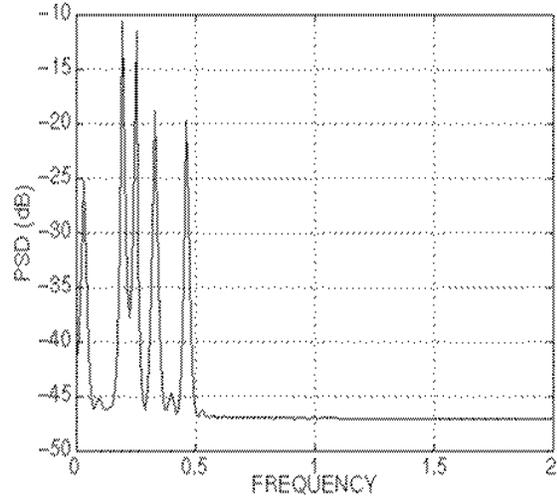}

}}
\caption[]{n.e. estimate.}
\end{figure}

We now present the result for whitening pre-processing on one synthetic
signal (figures 14-16). We compare this technique with the standard n.e.\ .
\par\noindent
Signal frequencies=$0.1, 0.15, 0.2, 0.25, 0.3 \quad 1/s $
\par\noindent
Neural network estimates with whitening: $0.1, 0.15, 0.2, 0.25, 0.3 \quad
1/s $
\par\noindent
Neural network estimates without whitening: $0.1, 0.15, 0.2, 0.25, 0.3 \quad
1/s $

From this and other experiments we saw that when we used the whitening in
our n.e. the results were worse and more time consuming than the ones
obtained using the standard n.e. (i.e. without whitening the signal). For
these reasons whitening is not a suitable technique to improve our n.e.\ .

\begin{figure}
\vbox{
\hbox{
\includegraphics[width=8cm , height=7cm]{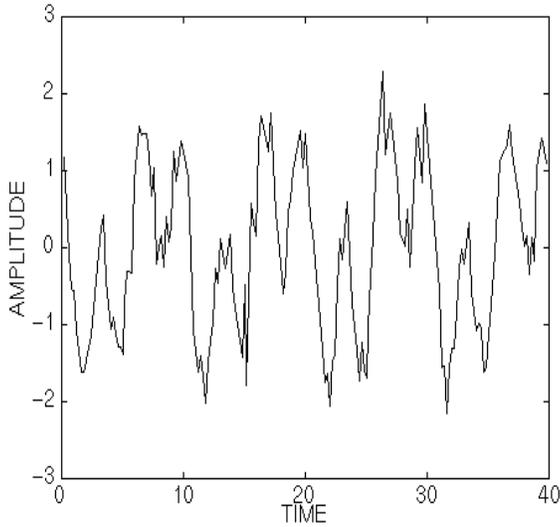}
}}
\caption[]{Synthetic Signal.}
\end{figure}

\begin{figure}
\vbox{
\hbox{
\includegraphics[width=8cm , height=7cm]{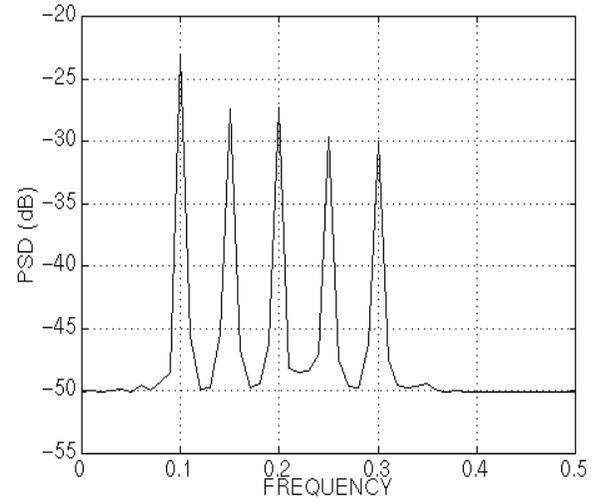}

}}
\caption[]{n.e. estimate without whitening.}
\end{figure}

\begin{figure}
\vbox{
\hbox{
\includegraphics[width=8cm , height=7cm]{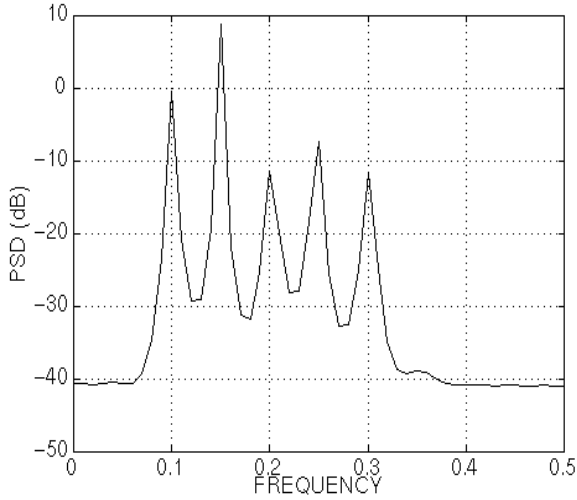}

}}
\caption[]{n.e. estimate with withening.}
\end{figure}

\subsection{ Comparison of the n.e. with the Lomb's Periodogram\label%
{realistic data}}

Here we present a set of synthetic signals generated by random varying the
noise variance, the phase and the deleted points with Montecarlo methods.
The signal is a sinusoid ($0.5\cos (2\pi 0.1t+\phi )+R(t)$) with frequency 
$0.1$ $Hz$, $R(t)$ the Gaussian random noise with $0$ mean composed by $100$
points, with a random phase. We follow Horne \& Baliunas (Horne \& Baliunas
1986) for the choice of the signals.

We generated two different series of samples depending on the number of
deleted points: the first one with $50$ deleted points, the second one with 
$80$ deleted points. We made 100 experiments for each variance value. The
results are shown in Table~\ref{table2} and Table~\ref{table3}, and compared 
with the Lomb's P
because it works better than the Scargle's P with unevenly spaced data,
introducing confidence intervals which are useful to identify the accepted
peaks.

The results show that both the techniques obtain a comparable performance.

\par\noindent
\begin{table*}[tbp]
\caption[]{Synthetic signal with 50 deleted points, 
frequency interval $\left[ \frac{2\pi}{T}, \frac{\pi N_{o}}{T} \right]$,
MSE = Mean Square Error,
$T$ = total period $(X_{max} - X_{min})$ and $N_{o}$ = total number of 
points.}
\label{table2}

\begin{flushleft}
\begin{tabular}{|c|c|c|c|c|c|c|c|}
\hline
  &         & \multicolumn{3}{|c|}{Lomb's P} & \multicolumn{2}{|c|}{n.e.} \\ 
\hline
Error Variance $\sigma^{2}$ & S.N.R. $\xi = \frac{X_{o}}{2\sigma^{2}}$ & 
Mean & Variance & MSE & Mean & Variance & MSE \\
\hline
0.75 & 0.2 & 0.1627 & 0.0140 & 0.0178 & 0.1472 & 0.0116 & 0.0131 \\
0.5 & 0.5 & 0.1036 & 0.0013 & 0.0013 & 0.1020 & 3.0630 e$^{-4}$ & 
3.0725 e$^{-4}$ \\
0.1 & 12.5 & 0.1000 & 1.0227 e$^{-8}$ & 1.0226 e$^{-8}$ & 0.1000 & 
6.1016 e$^{-8}$ & 6.2055 e$^{-8}$ \\
0.001 & 1250 & 0.1000 & 2.905 e$^{-9}$ & 2.3139 e$^{-9}$ & 0.1000 &
3.8130 e$^{-32}$ & 0.00000 \\
\hline
\end{tabular}
\end{flushleft}
\end{table*}

\par\noindent
\begin{table*}[tbp]
\caption[]{Synthetic signal with 80 deleted points, 
frequency interval $\left[ \frac{2\pi}{T}, \frac{\pi N_{o}}{T} \right]$,
MSE = Mean Square Error,
$T$ = total period $(X_{max} - X_{min})$ and $N_{o}$ = total number of 
points.}
\label{table3}
\begin{flushleft}
\begin{tabular}{|c|c|c|c|c|c|c|c|}
\hline
  &         & \multicolumn{3}{|c|}{Lomb's P} & \multicolumn{2}{|c|}{n.e.} \\ 
\hline
Error Variance $\sigma^{2}$ & S.N.R. $\xi = \frac{X_{o}}{2\sigma^{2}}$ & 
Mean & Variance & MSE & Mean & Variance & MSE \\
\hline
0.75 & 0.2  & 0.2323 & 0.0205 & 0.0378 & 0.2055 & 0.0228 & 0.0337 \\
0.5 & 0.5  & 0.2000 & 0.0190 & 0.0288 & 0.2034 & 0.0245 & 0.0349 \\
0.1  & 12.5 & 0.1000 & 2.37 e$^{-7}$   & 2.3648 e$^{-7}$ & 0.1004 & 
1.8437 e$^{-5}$ & 1.8435 e$^{-5}$ \\
0.001 & 1250 & 0.1000 & 8.6517 e$^{-8}$ & 8.5931 e$^{-8}$ & 0.1000 &
4.7259 e$^{-8}$ & 4.7259 e$^{-8}$ \\
\hline
\end{tabular}
\end{flushleft}
\end{table*}

\subsection{ Real data\label{real data}}

The first real signal is related to the Cepheid SU Cygni (Fernie 1979). The
sequence was obtained with the photometric tecnique UBVRI and the sampling
made from June to December 1977. The light curve is composed by 21 samples
in the V band, and a period of $3.8^{d}$, as shown in figure 17. In this
case, the parameters of the n.e. are: $N=10$, $p=2$, $\alpha =20$, $\mu
=0.001$. The estimate frequency interval is $\left[ 0(1/JD),0.5(1/JD)\right] 
$. The estimated frequency is $0.26$ (1/JD) in agreement with the Lomb's P,
but without showing any spurious peak (see figures 18 and 19).

\begin{figure}
\vbox{
\hbox{
\includegraphics[width=8cm , height=7cm]{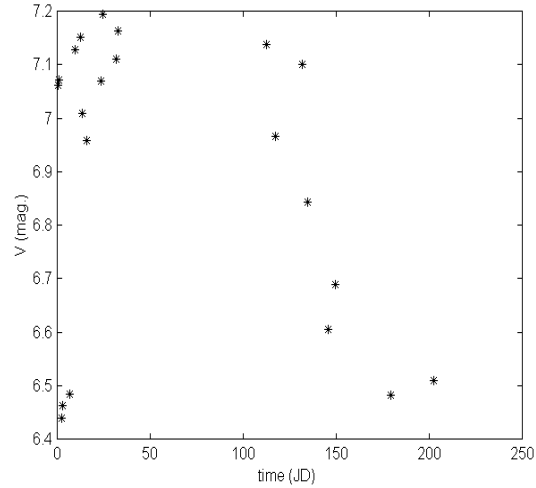}

}}
\caption[]{Light curve of SU Cygni.}
\end{figure}

\begin{figure}
\vbox{
\hbox{
\includegraphics[width=8cm , height=7cm]{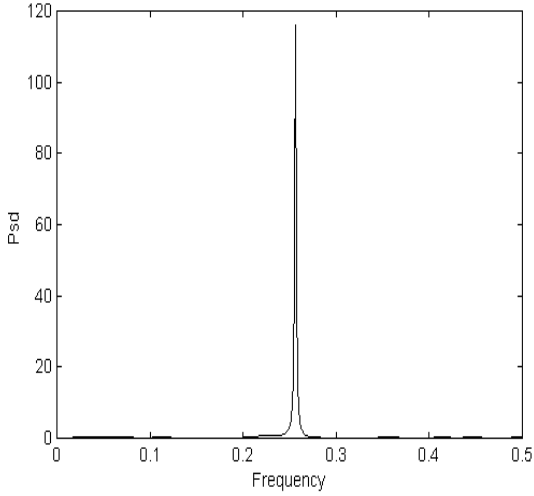}

}}
\caption[]{n.e. estimate of SU Cygni.}
\end{figure}

\begin{figure}
\vbox{
\hbox{
\includegraphics[width=8cm , height=7cm]{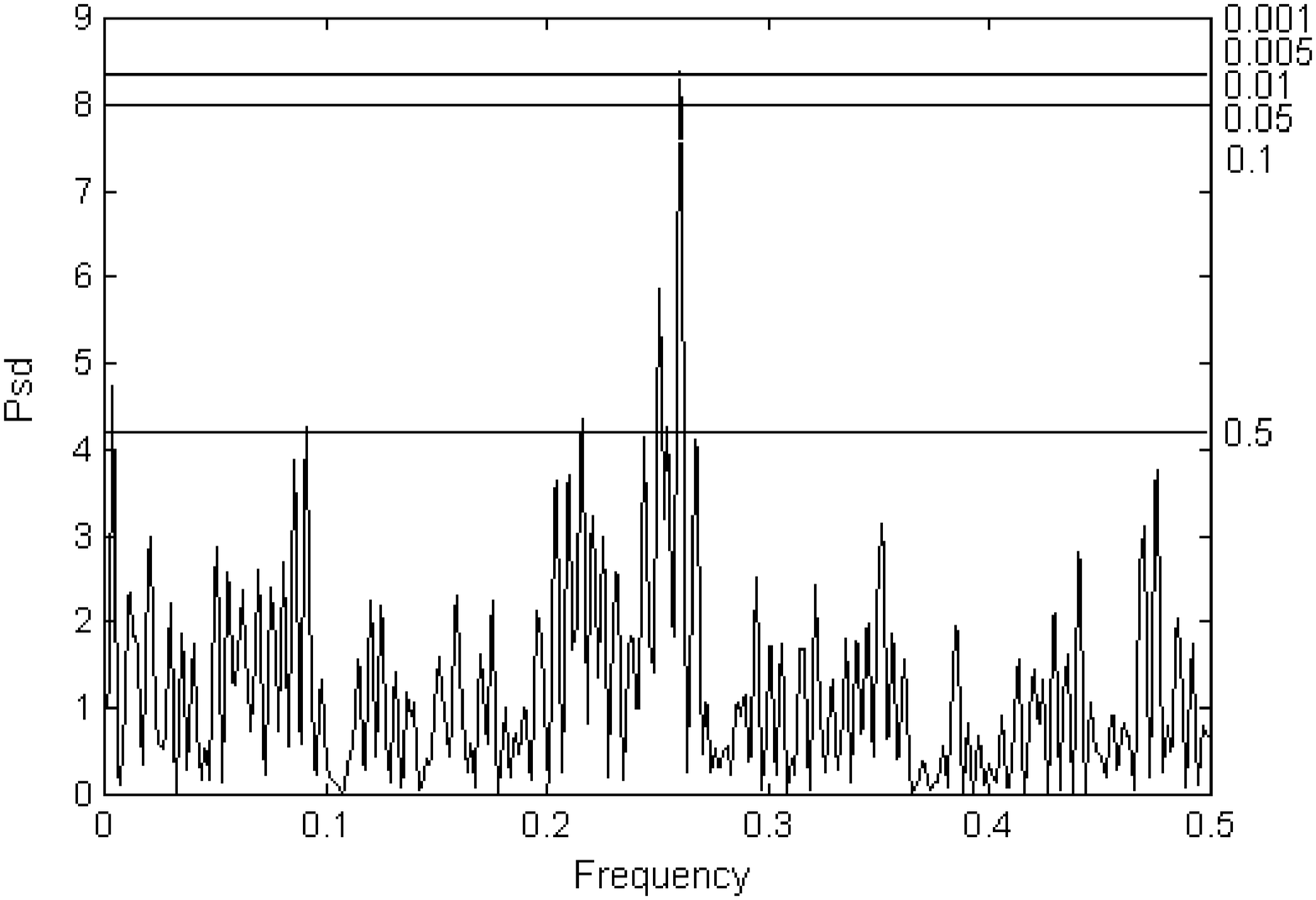}

}}
\caption[]{Lomb's P estimate of SU Cygni.}
\end{figure}

The second real signal is related to the Cepheid U Aql (Moffet and Barnes
1980). The sequence was obtained with the photometric tecnique BVRI and the
sampling made from April 1977 to December 1979. The light curve is composed
by 39 samples in the V band, and a period of $7.01^{d}$, as shown in figure
20. In this case, the parameters of the n.e. are: $N=20$, $p=2$, $\alpha =5$, 
$\mu =0.001$. The estimate frequency interval is $\left[ 0(1/JD),0.5(1/JD)
\right] $. The estimated frequency is $0.1425$ (1/JD) in agreement with the
Lomb's P, but without showing any spurious peak (see figures 21 and 22).

\begin{figure}
\vbox{
\hbox{
\includegraphics[width=8cm , height=7cm]{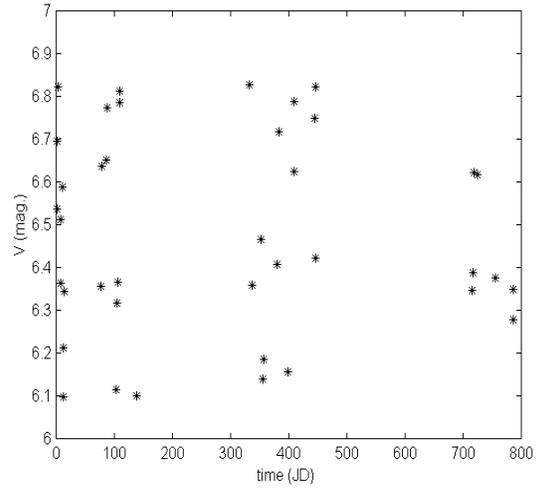}

}}
\caption[]{Light curve of U Aql.}
\end{figure}

\begin{figure}
\vbox{
\hbox{
\includegraphics[width=8cm , height=7cm]{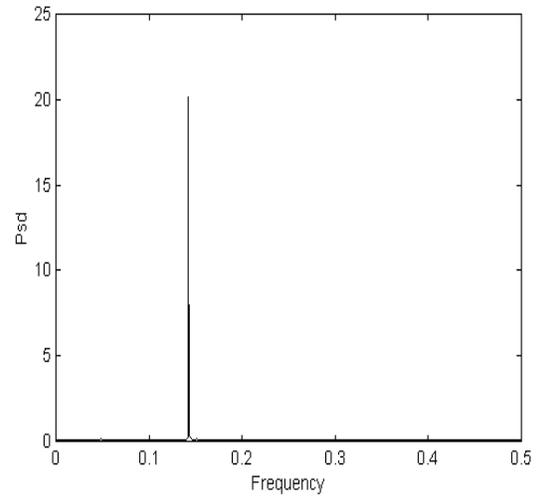}

}}
\caption[]{n.e. estimate of U Aql.}
\end{figure}

\begin{figure}
\vbox{
\hbox{
\includegraphics[width=8cm , height=7cm]{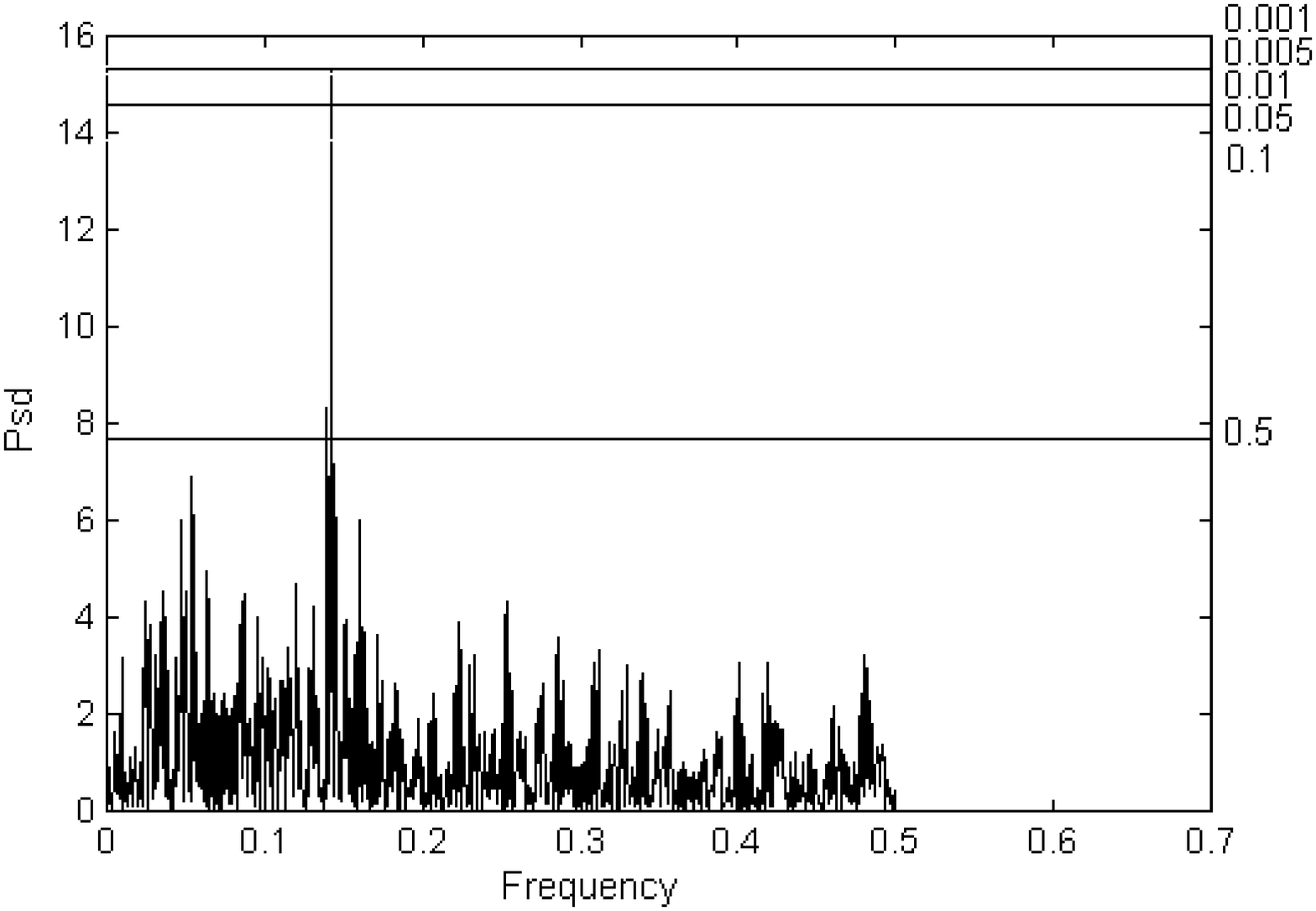}

}}
\caption[]{Lomb's P estimate of U Aql.}
\end{figure}

The third real signal is related to the Cepheid X Cygni (Moffet and Barnes
1980). The sequence was obtained with the photometric technique BVRI and the
sampling made from April 1977 to December 1979. The light curve is composed
by 120 samples in the V band, and a period of $16.38^{d}$, as shown in
figure 23. In this case, the parameters of the n.e. are: $N=70$, $p=2$, 
$\alpha =5$, $\mu =0.001$. The estimate frequency interval is $\left[
0(1/JD),0.5(1/JD)\right] $. The estimated frequency is $0.061$ (1/JD) in
agreement with the Lomb's P, but without showing any spurious peak (see
figures 24 and 25).

\begin{figure}
\vbox{
\hbox{
\includegraphics[width=8cm , height=7cm]{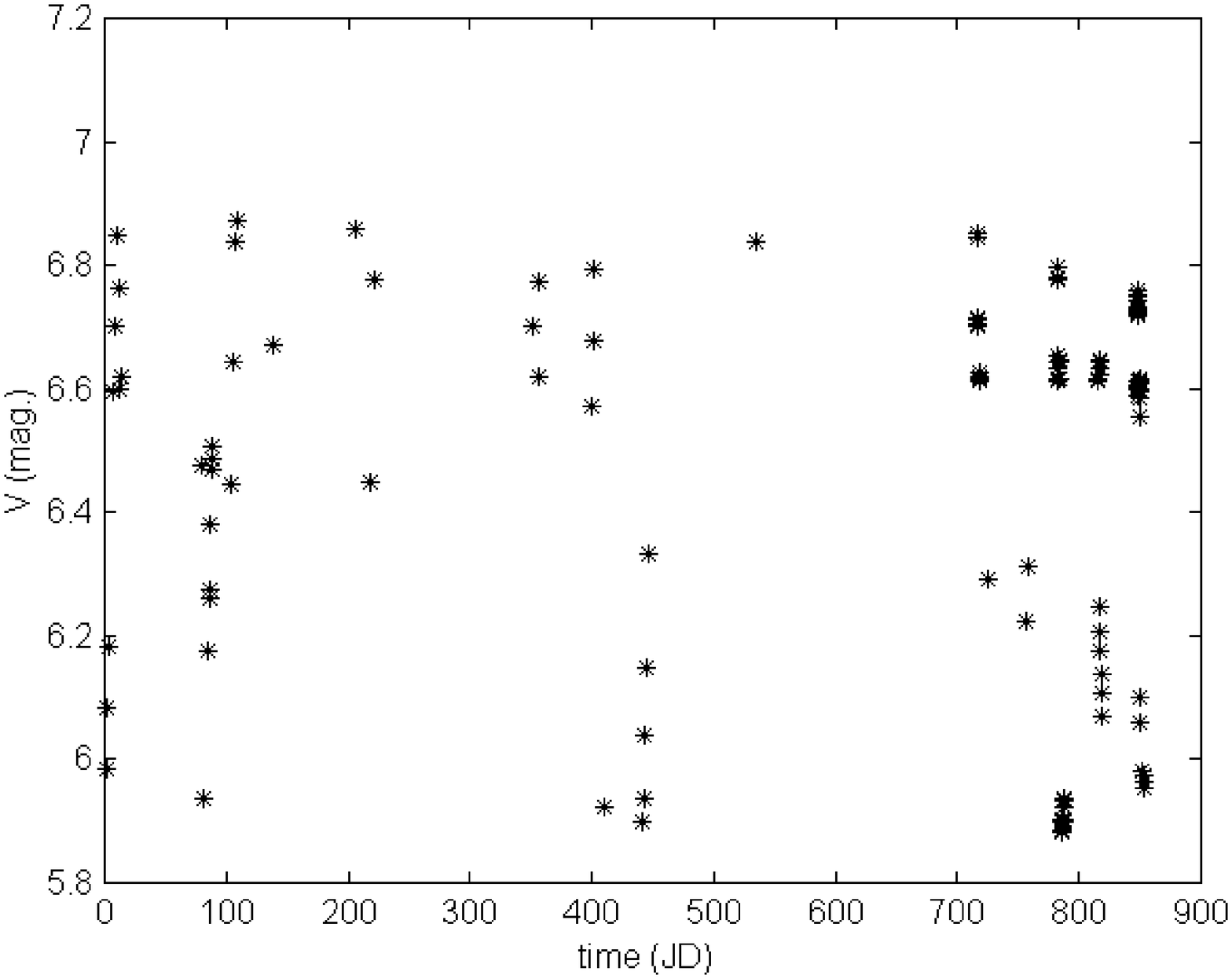}

}}
\caption[]{Light curve of X Cygni.}
\end{figure}

\begin{figure}
\vbox{
\hbox{
\includegraphics[width=8cm , height=7cm]{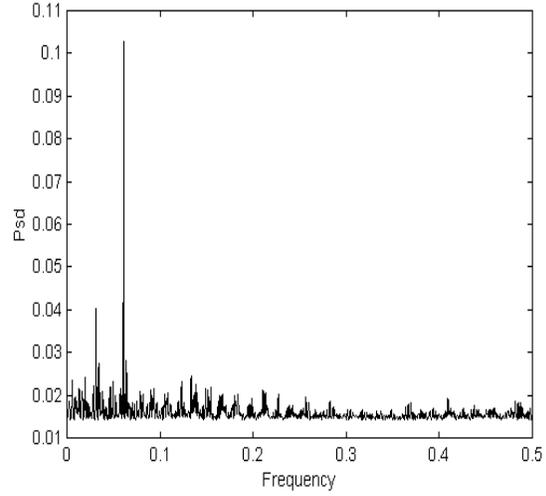}

}}
\caption[]{n.e. estimate of X Cygni.}
\end{figure}

\begin{figure}
\vbox{
\hbox{
\includegraphics[width=8cm , height=7cm]{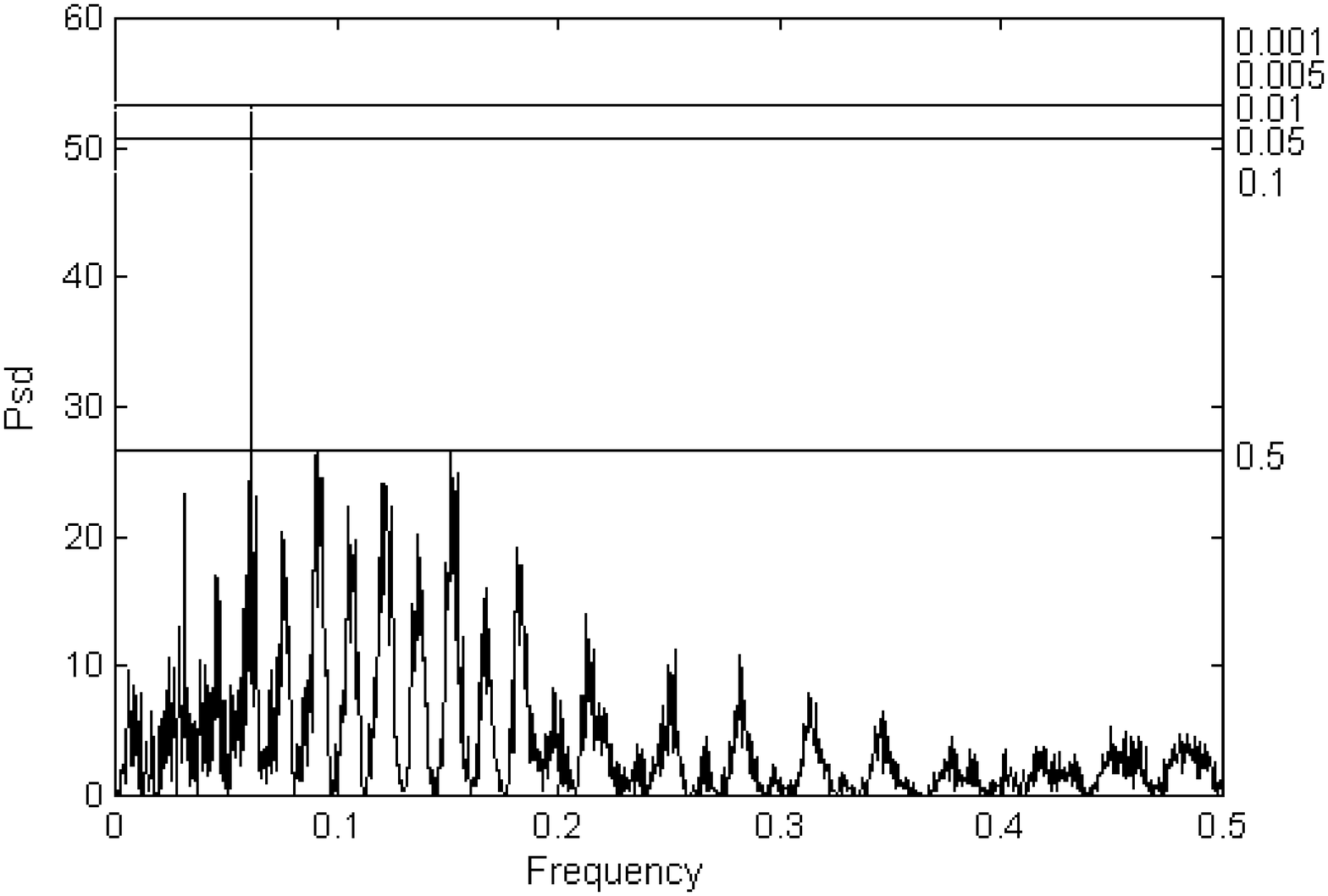}

}}
\caption[]{Lomb's P estimate of X Cygni.}
\end{figure}

The fourth real signal is related to the Cepheid T Mon (Moffet and Barnes
1980). The sequence was obtained with the photometric technique BVRI and the
sampling made from April 1977 to December 1979. The light curve is composed
by $24$ samples in the V band, and a period of $27.02^{d}$, as shown in
figure 26. In this case, the parameters of the n.e. are: $N=10$, $p=2$, 
$\alpha =5$, $\mu =0.001$. The estimate frequency interval is $\left[
0(1/JD),0.5(1/JD)\right] $. The estimated frequency is $0.037$ (1/JD) (see
figure 28).

\begin{figure}
\vbox{
\hbox{
\includegraphics[width=8cm , height=7cm]{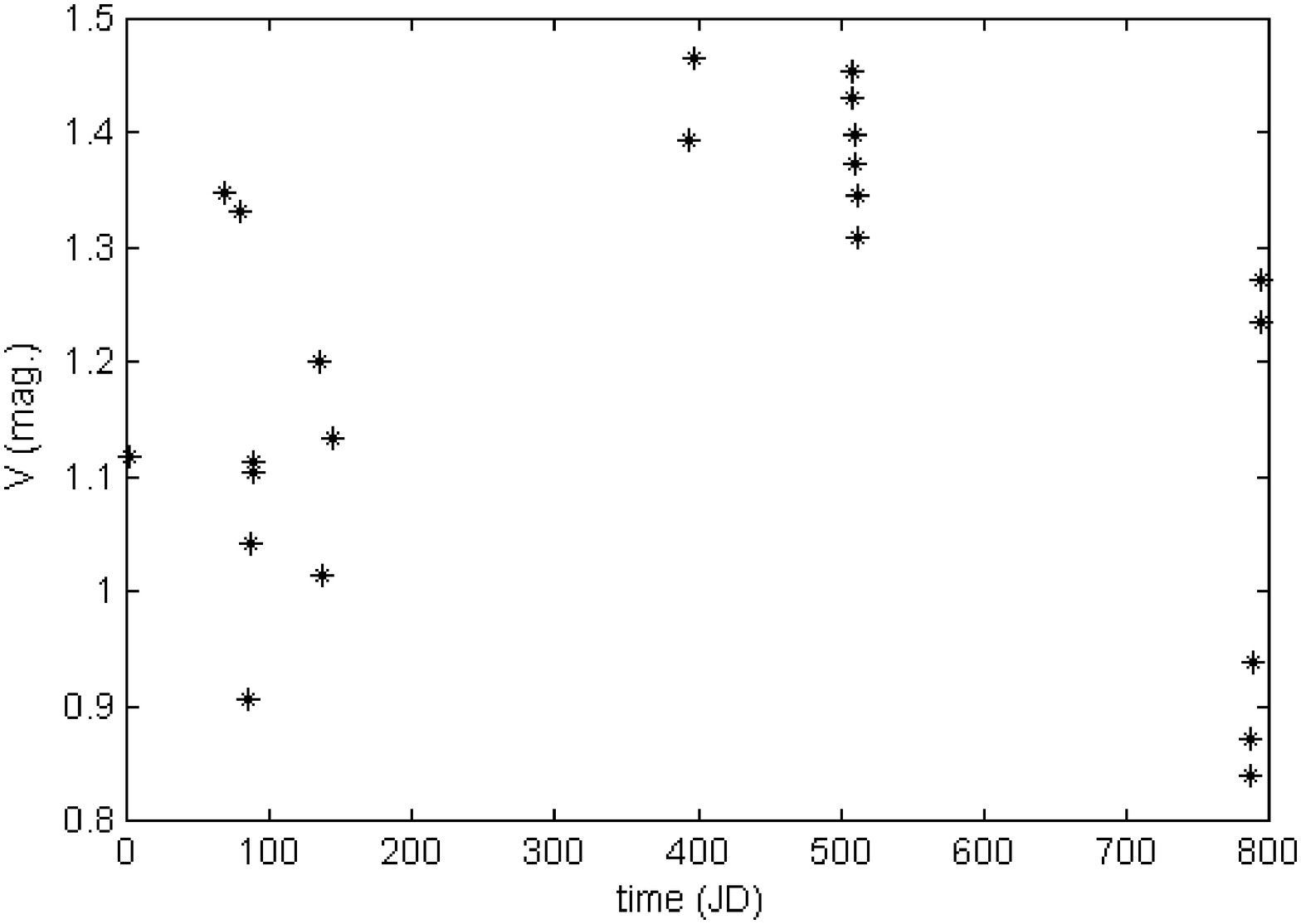}

}}
\caption[]{Light curve of T Mon.}
\end{figure}

\begin{figure}
\vbox{
\hbox{
\includegraphics[width=8cm , height=7cm]{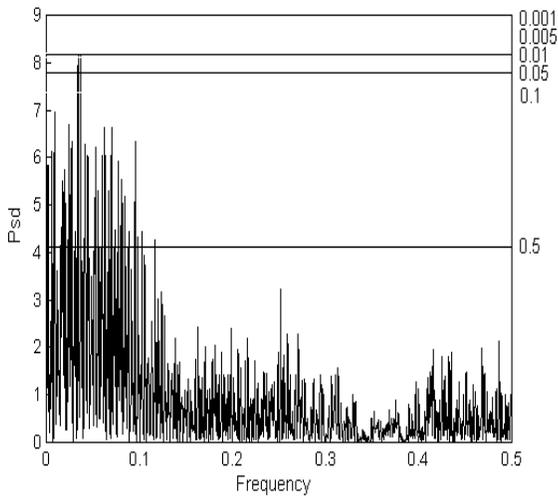}

}}
\caption[]{Lomb's P estimate of T Mon.}
\end{figure}

\begin{figure}
\vbox{
\hbox{
\includegraphics[width=8cm , height=7cm]{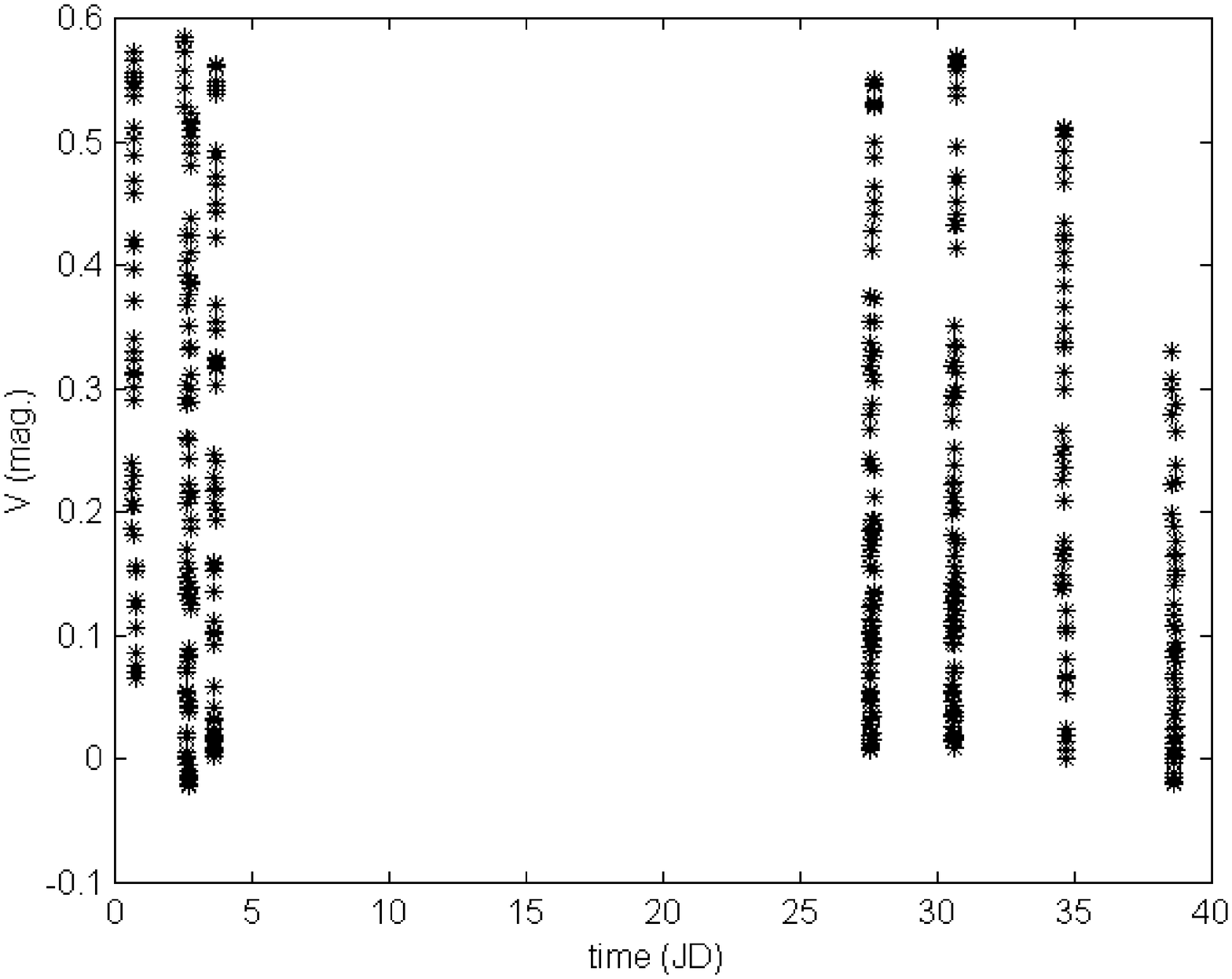}

}}
\caption[]{n.e. estimate of T Mon.}
\end{figure}

The Lomb's P does not work in this case because there many peaks, and at
least two greater than the threshold of the most accurate confidence
interval (see figure 27).

The fifth real signal we used for the test phase is a light curve in the
Johnson's system (Binnendijk 1960) for the eclipsing binary U Peg (see
figure 29 and 30). This system was observed photoelectrically in the
effective wavelengths 5300 A and 4420 A with the 28-inch reflecting
telescope of the Flower and Cook Observatory during October and November,
1958.

\begin{figure}
\vbox{
\hbox{
\includegraphics[width=8cm , height=7cm]{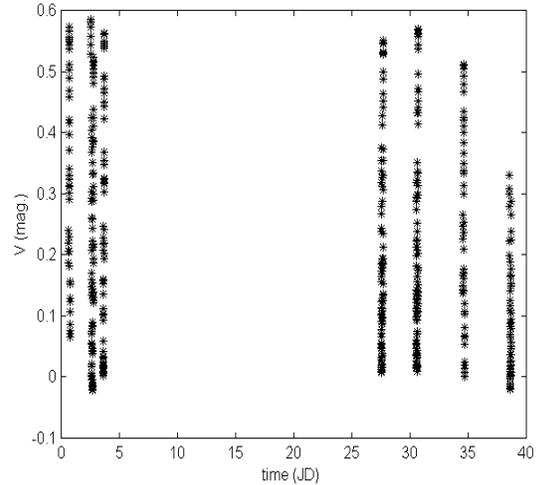}

}}
\caption[]{Light curve of U Peg.}
\end{figure}

\begin{figure}
\vbox{
\hbox{
\includegraphics[width=8cm , height=7cm]{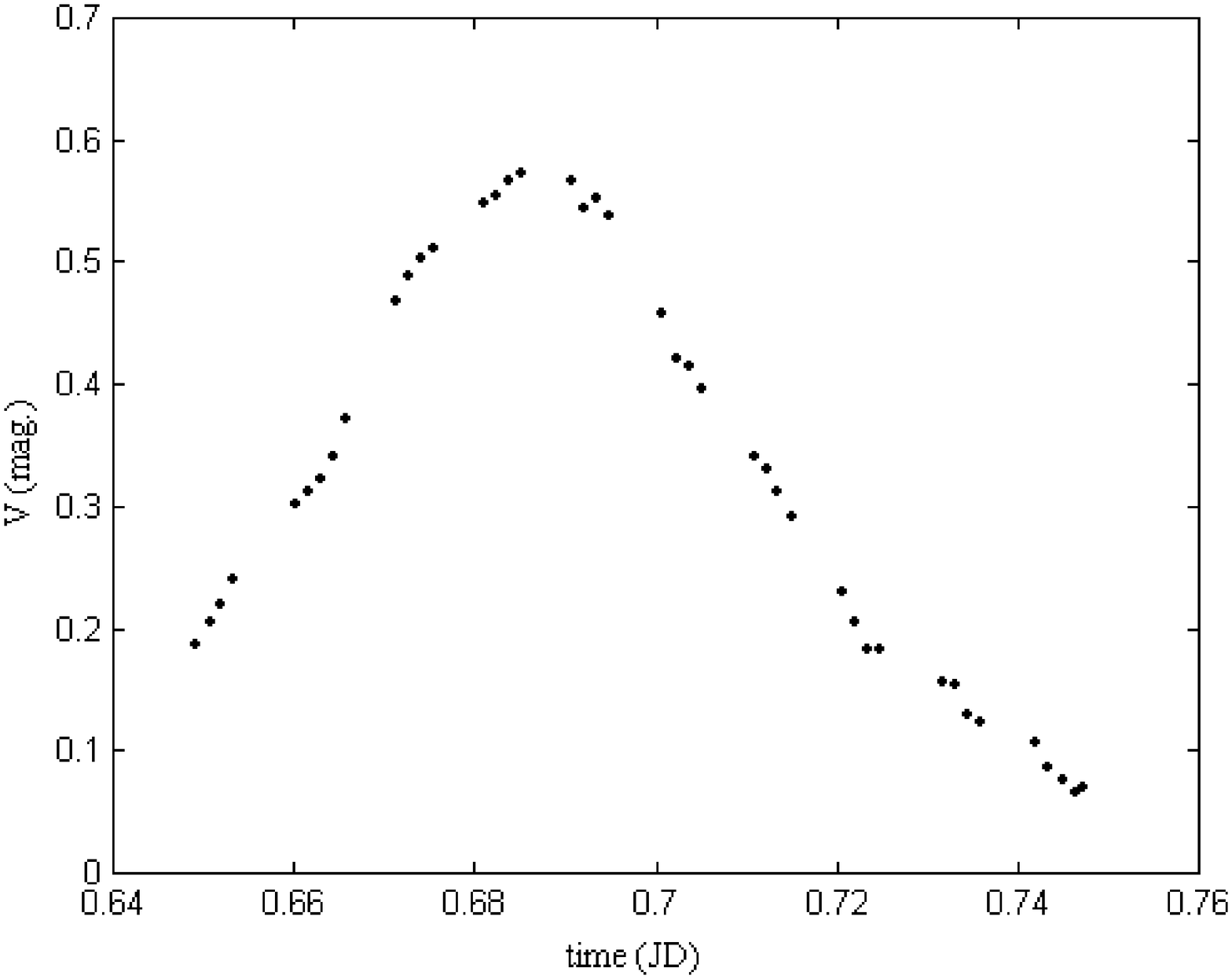}

}}
\caption[]{Light curve of U Peg (first window).}
\end{figure}

We made several experiments with the n.e., and we elicited a dependence of
the frequency estimate on the variation of the number of elements for input
pattern. The optimal experimental parameters for the n.e. are: $N=300$, $%
\alpha =5$; $\mu =0.001$. The period found by the n.e. is expressed in $JD$
and is not in agreement with results cited in literature (Binnendijk 1960),
(Rigterink 1972), (Zhai et al. 1984),(Lu 1985) and (Zhai et al. 1988). The
fundamental frequency is $5.4$ $1/JD$ (see figure 31) instead of $2.7$ 
$1/JD$. We obtain a frequency double of the observed one. Lomb's P has some 
high peaks as in the previous experiments and the estimated frequency is 
always the double of the observed one (see figure 32).

\begin{figure}
\vbox{
\hbox{
\includegraphics[width=8cm , height=7cm]{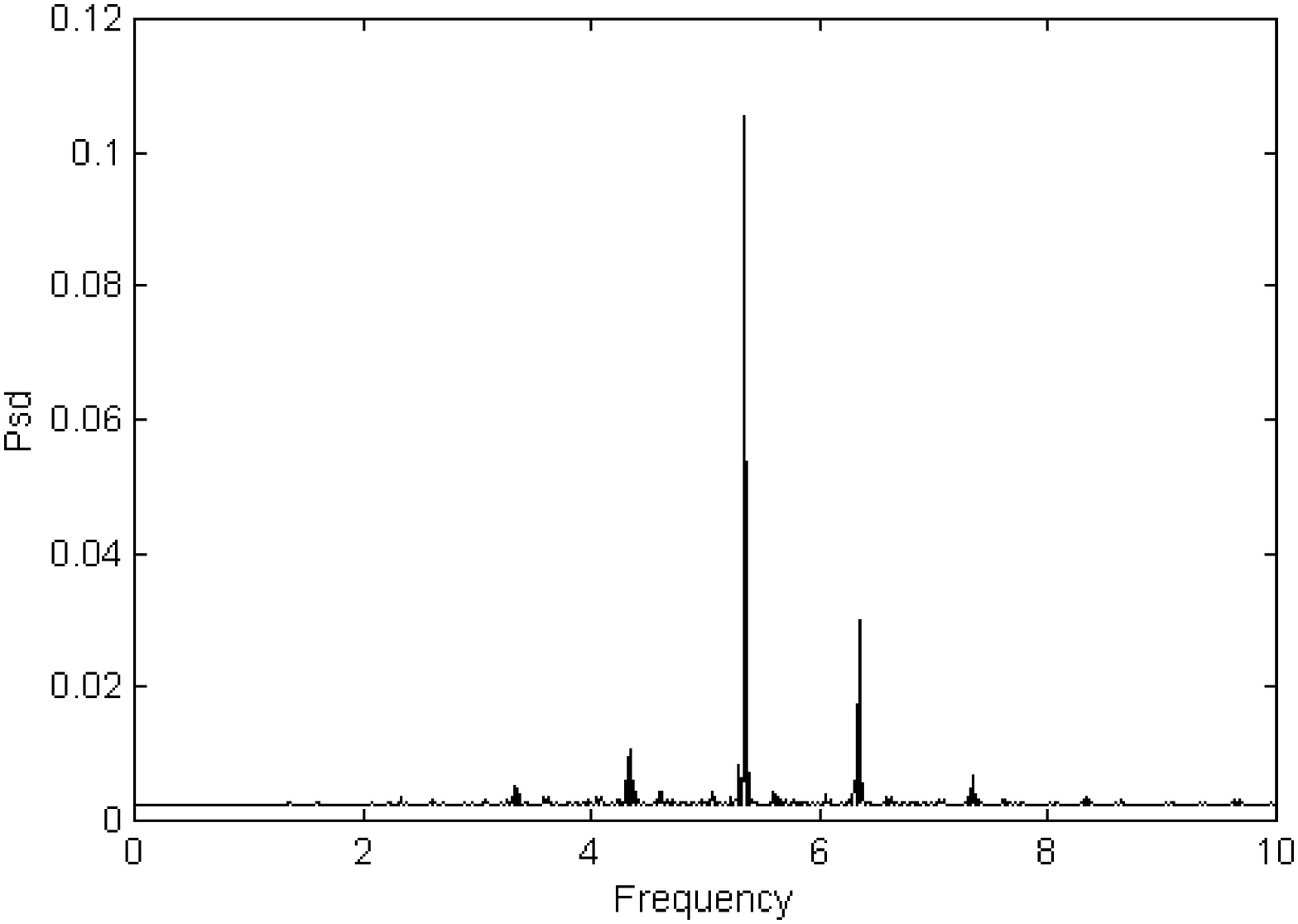}

}}
\caption[]{n.e. estimate of U Peg.}
\end{figure}

\begin{figure}
\vbox{
\hbox{
\includegraphics[width=8cm , height=7cm]{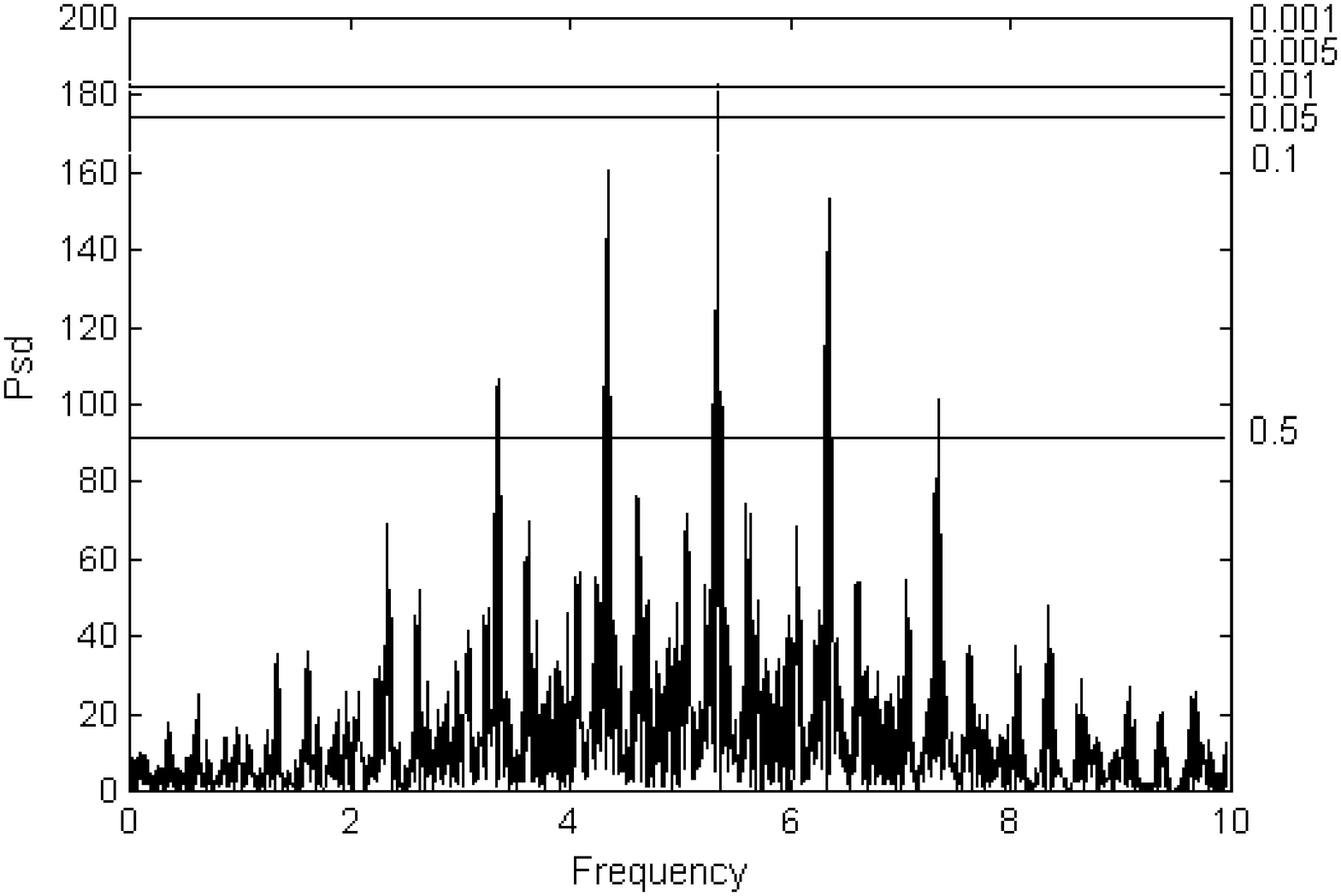}

}}
\caption[]{Lomb's P of U Peg.}
\end{figure}

\section{Conclusions}

We have realised and experimented a new method for spectral analysis for
unevenly sampled signals based on three phases: preprocessing, extraction of
principal eigenvectors and estimate of signal frequencies. This is done,
respectively, by input normalization, nonlinear PCA neural networks, and the
Multiple Signal Classificator algorithm. First of all, we have shown that
neural networks are a valid tool for spectral analysis.

However, above all, what is really important is that neural networks, as
realised in our neural estimator system, represent a new tool to face and
solve a problem tied with data acquisition in many scientific fields: the
unevenly sampling scheme.

Experimental results have shown the validity of our method as an alternative
to Periodogram, and in general to classical spectral analysis, mainly in
presence of few input data, few a priori information and high error
probability. Moreover, for unevenly sampled data, our system offers great
advantages with respect to P. First of all, it allows us to use a simple and
direct way to solve the problem as shown in all the experiments with
synthetic and Cepheid's real signals. Secondly, it is insensitive to the
frequency interval: for example, if we expand our interval in the SU Cygni
light curve, while our system finds the correct frequency, the Lomb's P
finds many spurious frequencies, some of them greater than the confidence
threshold, as shown in figures 33 and 34.

\begin{figure}
\vbox{
\hbox{
\includegraphics[width=8cm , height=7cm]{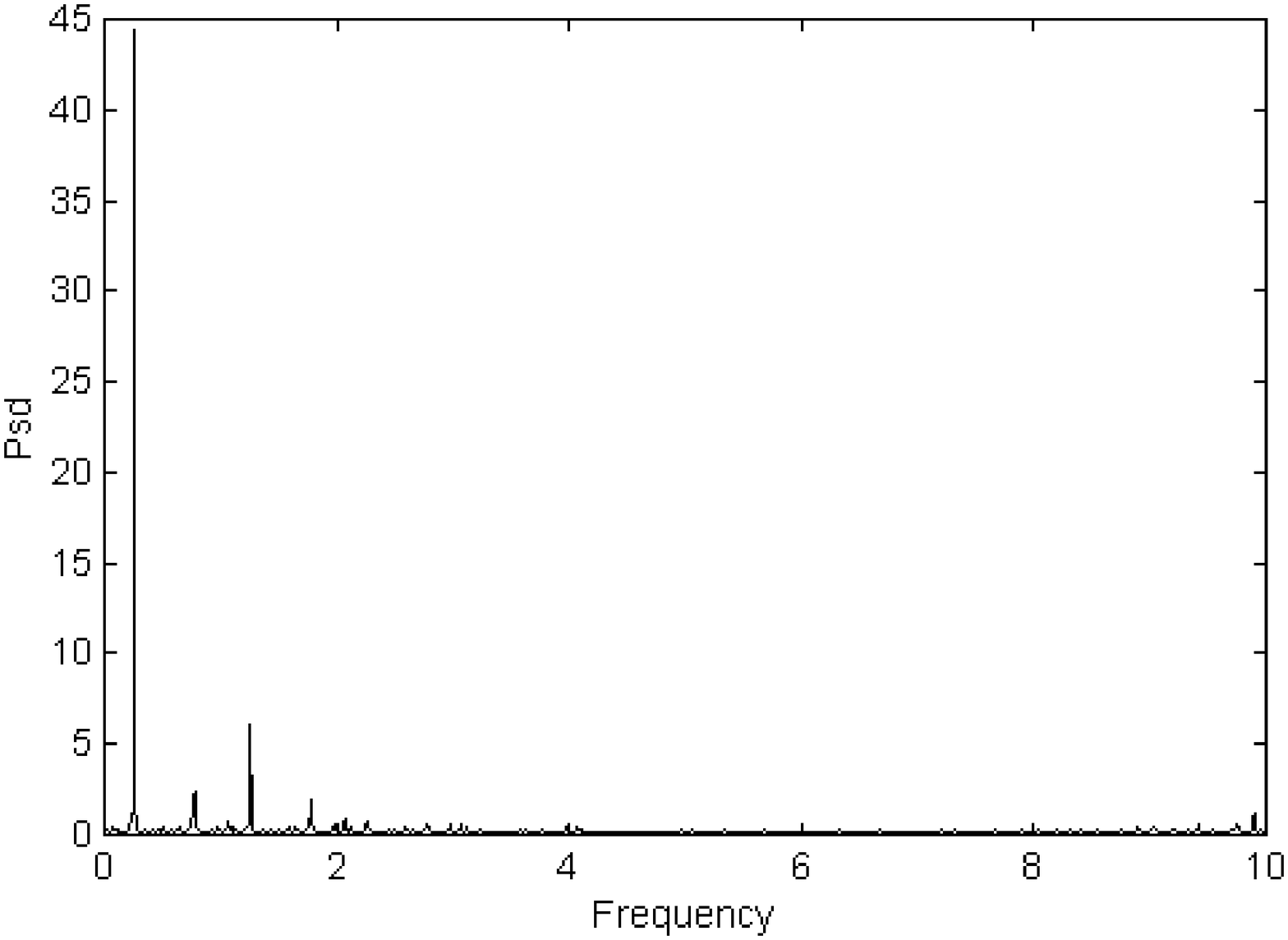}

}}
\caption[]{n.e. estimate of SU Cygni with enlarged window.}
\end{figure}

\begin{figure}
\vbox{
\hbox{
\includegraphics[width=8cm , height=7cm]{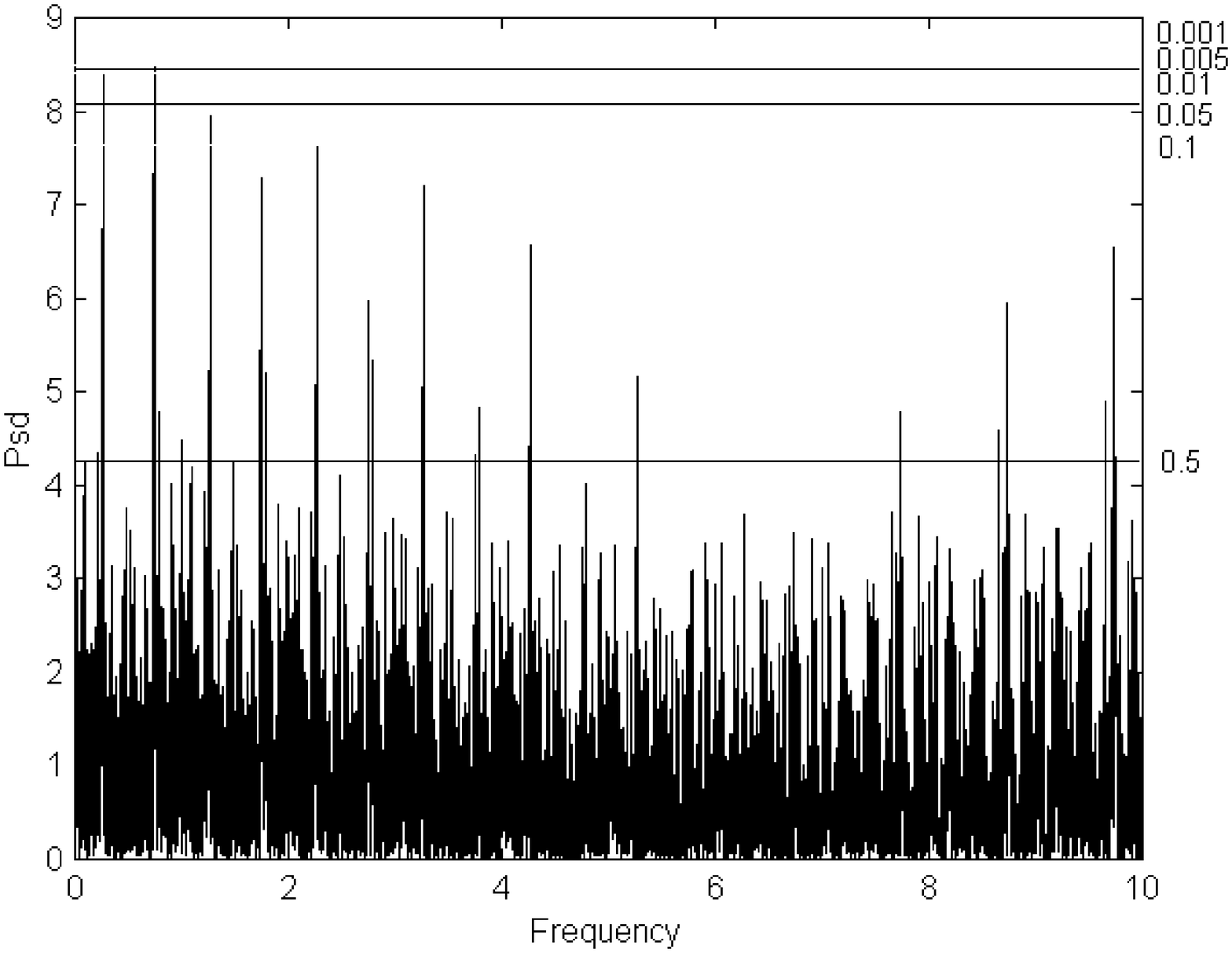}

}}
\caption[]{Lomb's P of SU Cygni with enlarged window.}
\end{figure}

Furthermore, when we have a multifrequency signal, we can use our system
also if we do not know the frequency number. In fact, we can detect one
frequency at each time and continue the processing after the cancellation of
the detected periodicity by IIR filtering.

A point worth of noting is the failure to find the right frequency in the
case of eclipsing binary for both our method and Lomb's P. Taking account of
the morphology of eclipsing light curve with two minima, this fact can not
be of concern because in practical cases the important thing is to have a
first guess of the orbital frequency. Further refinement will be possible
through a wise planning of observations. In any case we have under study
this point to try to find a method to solve the problem.

\section*{Acknowledgments}
The authors would like to thank Dr. M. Rasile for the experiments related to the
model selection and an unknown referee for his comments who helped the
authors to improve the paper.

\end{document}